\documentclass[useAMS,usenatbib]{mn2e}

\usepackage{amsmath,amssymb,mathrsfs,graphicx,float,latexsym,url}
\usepackage{epstopdf,ragged2e}
\usepackage{natbib,url}
\usepackage{hyperref}

\usepackage[usenames,dvipsnames]{color}
\definecolor{darkred}{rgb}{0.5,0,0}
\definecolor{darkgreen}{rgb}{0,0.5,0}
\definecolor{darkblue}{rgb}{0,0,0.5}
\definecolor{prussian}{rgb}{0.0, 0.19, 0.33}
\definecolor{richelectricblue}{rgb}{0.03, 0.57, 0.82}
\definecolor{teal}{rgb}{0.0, 0.5, 0.5}
\definecolor{mediumseagreen}{rgb}{0.24, 0.7, 0.44}
\definecolor{lust}{rgb}{0.9, 0.13, 0.13}
\definecolor{ballblue}{rgb}{0.13, 0.67, 0.8}
\definecolor{darkcyan}{rgb}{0.0, 0.55, 0.55}
\definecolor{mountainmeadow}{rgb}{0.19, 0.73, 0.56}
\definecolor{palecarmine}{rgb}{0.69, 0.25, 0.21}
\definecolor{richcarmine}{rgb}{0.84, 0.0, 0.25}
\definecolor{tangelo}{rgb}{0.98, 0.3, 0.0}
\definecolor{venetian}{rgb}{0.784,0.031,0.082}
\definecolor{bdfrance}{rgb}{0.192,0.549,0.906}
\usepackage{mathtools}
\usepackage{amsbsy}
\usepackage{bm}
\usepackage{float}

\newcommand{\tO}{\tau_{\text{Ohm}}}
\newcommand{\tH}{\tau_{\text{Hall}}}
\newcommand{\vPL}{v_{\text{pl}}}
\newcommand{\bb}{\boldsymbol{B}}
\newcommand{\bvPL}{\boldsymbol{v}_{\text{pl}}}

\newcommand{\Rs}{R_{\star}}

\newcommand{\rLC}{R_{\rm LC}}

\def\apj{{ApJ}}
\def\aj{{AJ}}
\def\apjs{{The Astrophysical Journal Supplement}}
\def\apjl{{ApJL}}

\def\aap{{A\&A}}

\def\mnras{{MNRAS}}

\def\pasj{{Publications of the Astronomical Society of Japan}}

\def\prd{{Physical Review D}}
\def\prl{{Phys. Rev. Lett.}}

\def\prc{{Physical Review C}}

\def\04a{{2004 a}}
\def\04b{{2004 b}}

\begin{document}

\title[A magnetar interpretation for GLEAM-X J1627]
{Evolutionary implications of a magnetar interpretation for GLEAM-X J162759.5--523504.3}
\author[A. G.~Suvorov \& A.~Melatos]{Arthur G. Suvorov$^{1,2}$\thanks{arthur.suvorov@manlyastrophysics.org} 
and Andrew Melatos$^{3,4}$\\
$^1$Manly Astrophysics, 15/41-42 East Esplanade, Manly, NSW 2095, Australia\\
$^2$Theoretical Astrophysics, Eberhard Karls University of T{\"u}bingen, T{\"u}bingen, D-72076, Germany\\
$^3$School of Physics, University of Melbourne, Parkville VIC 3010, Australia\\
$^4$ARC Centre of Excellence for Gravitational Wave Discovery (OzGrav), Hawthorn VIC 3122, Australia
}

\date{Accepted ?. Received ?; in original form ?}

\pagerange{\pageref{firstpage}--\pageref{lastpage}} \pubyear{?}

\maketitle
\label{firstpage}

\begin{abstract}

\noindent The radio pulsar GLEAM-X J162759.5--523504.3 has an extremely long spin period ($P = 1091.17\, \mbox{s}$), and yet {seemingly} continues to spin down rapidly ($\dot{P} < 1.2 \times 10^{-9}\, \mbox{ss}^{-1}$). The magnetic field strength that is implied, if the source is a neutron star undergoing magnetic dipole braking, {could} exceed $10^{16}\,\mbox{G}$. This object may therefore be the most magnetised neutron star observed to date. In this paper, a critical analysis of a magnetar interpretation for the source is provided. (i) A minimum polar magnetic field strength of $B \sim 5 \times 10^{15}\,\mbox{G}$ appears to be necessary for the star to activate as a radio pulsar, based on conventional `death valley' assumptions. (ii) Back-extrapolation from magnetic braking and Hall-plastic-Ohm decay {suggests} that a large angular momentum reservoir was available at birth to support intense field amplification. (iii) The observational absence of X-rays constrains the star's field strength and age, as the competition between heating from field decay and Urca cooling implies a surface luminosity as a function of time. If the object is an isolated, young ($\sim 10\, \mbox{kyr}$) magnetar with a present-day field strength of $B \gtrsim 10^{16}\,\mbox{G}$, {the upper limit ($\approx 10^{30}\, \mbox{erg s}^{-1}$)} set on its thermal luminosity suggests it is cooling via a direct Urca mechanism.
\end{abstract}



\section{Introduction} 
\label{sec:intro}

\cite{hw22} recently reported that observations, made between January and March of 2018 with the Murchison Widefield Array (MWA) in the $72-231\, \mbox{MHz}$ band, revealed the presence of a pulsating Galactic source, since named GLEAM-X J162759.5--523504.3 (henceforth GLEAM-X J1627), which is $88 \pm 1\%$ linearly polarised. Barycentric correction and alignment of the pulses established a periodicity with a pulsar-like regularity, $P = 1091.1690(5)\, \mbox{s}$, and further that the source is slowing down at a best-fit rate of $\dot{P} = 6 \times 10^{-10}\, \mbox{s s}^{-1}$. {(Though we note the data can only confidently assert that $|\dot{P}| < 1.2 \times 10^{-9}\, \mbox{ss}^{-1}$)}. The characteristic (polar) magnetic field strength relevant for a neutron star, $B \approx 6.4 \times 10^{19} \sqrt{P \dot{P}}\,\mbox{G}$ \cite[e.g.,][]{rs75}, is {arguably} in excess of $10^{16}$G. Furthermore, since the pulse structure of the source varies on $\sim$hour-long timescales in a way that is similar to what is seen from known radio magnetars, \cite{hw22} offered the tantalising conclusion that the source is an ultra-long period magnetar \cite[see also][]{hw22b,eks22,genc22,kat22}. This would likely make GLEAM-X J1627 the most magnetised neutron star observed to date\footnote{A list of known magnetars and their properties is maintained at \url{http://www.physics.mcgill.ca/~pulsar/magnetar/main.html}{; see also the Magnetar Outburst Online Catalogue \url{http://magnetars.ice.csic.es}.}} \citep{mcg14,coti18}.

Followup searches by the Swift X-ray Telescope (XRT) found no evidence for thermal or {soft} X-rays \citep{hw22}. Upper limits to the photon count were placed which, depending on the spectral fit, imply an upper limit to the flux. The strongest limit is placed at $1.9 \times 10^{-13}\, \mbox{erg s}^{-1} \mbox{cm}^{-2}$ for the absorbed flux in the 0.3--10keV band, with a marginally lower value ($\approx 1.5 \times 10^{-13}\, \mbox{erg s}^{-1} \mbox{cm}^{-2}$) applying for a blackbody fit at $kT \sim 0.1$~keV. Based on the greatest distance allowed by the dispersion measure, $d_{\text{max}} = 1.8\, \mbox{kpc}$, this gives $L_{\rm X} \leq 7 \times 10^{31}\, \mbox{erg s}^{-1}$ \citep{hw22}. {A deeper search using the Chandra X-ray Observatory was carried out by \cite{rea22}, who placed the even stricter upper limit $L_{\rm X} \leq 2 \times 10^{30}\, \mbox{erg s}^{-1}$, with the exact bound depending on assumptions about the spectral shape.} The latter upper limit would generally be expected of persistent, thermal emissions from a $\gtrsim$ Myr-old magnetar \citep{td96,tur11,mcg14}. Should GLEAM-X J1627 be a magnetar, its existence as a radio, but not an X-ray, source has a number of interesting implications for emission physics and magnetic field evolution in neutron stars, some of which we explore in this work.

It is generally put forth that electron-positron pair production, likely occurring in magnetospheric `gaps', is a necessary ingredient to spark the radio emissions seen from pulsars \citep{gj69,stur71,rs75} \cite[though cf.][]{melrose21}. Depending on the topological properties of the stellar magnetic field, a variety of different `death lines', defined through the threshold to generate the requisite pairs, comprise an overall `death valley' \citep{cr93,ha01}. Requiring that GLEAM-X J1627 reside outside of the valley allows us to assess the validity of a number of evolutionary scenarios. Having some idea about what surface field structures are permissible for the object `today', we can back-extrapolate from analytic or numerical simulations of Hall-plastic-Ohm decay in stellar crusts \citep{lg19,g22,koj22} to see what magnetic conditions at birth are {indicated}.

An intense magnetic field within the stellar core is expected to lead to ambipolar heating \citep{gr92,ag08}, as charged constituents (e.g., protons)  experience Lorentz forces that the uncharged components (e.g., neutrons) do not, leading to a kind of collisional friction that gradually heats up the star at the expense of the magnetic energy. From models of core-crust heat transfer \citep{pot03,tur11,vig13,anz22}, we can estimate the surface luminosity implied by the competition between ambipolar diffusion, mechanical dissipation, Joule heating, and particle backflow against neutrino cooling \citep{belo16}. The absence of thermal X-rays may then hint at an upper limit to the magnetic field strength, which can be compared with the requirements set by the radio activation.

In this paper, we revisit the magnetar interpretation of GLEAM-X J1627 in the context of death valley physics (Sec. \ref{sec:deathval}), Hall-plastic-Ohm evolutions (Sec. \ref{sec:hall}), braking mechanisms (Sec. \ref{sec:brake}), field amplifications at birth (Sec. \ref{sec:birth}), and ambipolar heating (Sec. \ref{sec:xrays}). {The conclusions are summarised in Sec. \ref{sec:conc}, emphasising that they depend on details of the model and cannot be asserted strongly based on the limited data at hand}. Quantities written with numerical subscripts are logarithmically normalised, e.g., $B_{16} = B/10^{16}$ for field strength $B$ measured in G.

\section{Magnetar nature of GLEAM-X J1627 and radio observations}
\label{sec:sec2}

Between January and March 2018\footnote{\cite{hw22} note that while the MWA, over the course of its eight years of operation, has accumulated around $\lesssim 200\, \mbox{hours}$ of observing time within 15$^{\circ}$ of GLEAM-X J1627, the data span many different array configurations, frequencies, and observing modes. It is therefore difficult to formally deduce the source duty cycle. Uncertainties notwithstanding, they argue that a $\sim 2\%$ duty cycle is likely, similar to the $\sim 5\%$ cycle of the radio-loud magnetar XTE J1810--197 \citep{eie21}.}, 71 pulses from GLEAM-X J1627 were detected by the MWA which, after alignment and barycentric correction, revealed a Galactic source pulsing with period $P = 1091.17\, \mbox{s}$. The best-fit value for the period derivative is $\dot{P} = 6 \times 10^{-10}\, \mbox{s s}^{-1}$, though \cite{hw22} noted that their analysis cannot exclude even larger values $\dot{P} < 1.2 \times 10^{-9}\, \mbox{s s}^{-1}$. For a neutron star moment of inertia $I_{0} \sim 10^{45}\, \mbox{g cm}^{2}$, the spin-down luminosity associated with the object is then $\dot{E}_{\text{sd}} \approx 4 \pi^2 I_{0} \dot{P} / P^3 \sim 10^{28}\, \mbox{erg s}^{-1}$, which is several orders of magnitude lower than the observed radio luminosity $L_{\nu} \approx 4 \times 10^{31}\, \mbox{erg s}^{-1}$, estimated assuming a best-fit distance of $d = 1.3 \pm 0.5\, \mbox{kpc}$. This puzzling feature, which is unique to GLEAM-X J1627, has prompted interest in a white dwarf interpretation for the source, essentially to boost $I_{0}$ \citep{loeb22,kat22}. However, \cite{erk22} argue that the beaming angle assumptions made by \cite{hw22} may be inappropriate for an object with such a long spin period \cite[see also][]{sz14,sz15}, and the radio luminosity may in fact be closer to $\approx 3 \times 10^{26}\, \mbox{erg s}^{-1}$ -- much lower than $\dot{E}_{\text{sd}}$.

In the standard picture of magnetic-dipole braking (cf. Sec. \ref{sec:brake}), the characteristic magnetic field strength for a neutron star reads $B_{p} \approx 6.4 \times 10^{19} \sqrt{P \dot{P}}\,\mbox{G}$. The $P$-$\dot{P}$ {upper limits} therefore {hint towards} a huge magnetic energy. This observation, combined with the high degree of linear polarisation in the pulses, led \cite{hw22} to suggest that GLEAM-X J1627 may be a magnetar. In this Section, we examine this suggestion in the theoretical context of radio emission mechanisms (Sec. \ref{sec:deathval}), crustal magnetic field evolution (Sec. \ref{sec:hall}), braking physics (Sec. \ref{sec:brake}), and numerical simulations of birth properties (Sec. \ref{sec:birth}). The Swift XRT observations of GLEAM-X J1627 are then discussed in Sec. \ref{sec:xrays}.

\subsection{Death valley}
\label{sec:deathval}

A neutron star crust provides a reservoir of free electrons that are continuously accelerated into the magnetosphere by induction-generated electric fields as the star spins. In `gap' regions where the \cite{gj69} charge density is comparatively low, the electric field along magnetic field lines can be sufficiently intense that photons emitted by accelerated charges possess the requisite energy to pair produce. It is generally put forth that $\rm e^{+} \rm e^{-}$ production is an essential ingredient in the powering of coherent radio emissions from neutron stars \citep{stur71,rs75,cr93,ha01} \cite[cf.][]{melrose21}. Secondary charges generated by curvature- or inverse-Compton-produced photons can themselves be accelerated and emit photons\footnote{Above hot polar caps with temperatures exceeding $\sim 10^6\, {\rm K}$, collisional interactions between photons could potentially also produce abundant pairs through the \cite{bw34} interaction $\gamma \gamma \rightarrow \rm e^{+} + \rm e^{-}$ \citep{jones22}.}, culminating in a pair cascade. Beam instabilities, where free energy associated with streaming motions is transferred to plasma waves, then lead to radio emission \cite[though again cf.][who argue this picture requires revision]{melrose21}.

In this scenario, there is a maximum potential drop, $\Delta V_{\text{max}}$, which can be produced in the magnetosphere, viz.
\begin{equation}
\Delta V_{\text{max}} \approx \frac {2 \pi^2 B_{d} \Rs^3} {c^2 P^2},
\end{equation}
for stellar radius $\Rs$, speed of light $c$, and polar \emph{dipole} strength $B_{d}$. This maximum must exceed that which is required for the pair production mechanism to activate. In a curvature radiation and polar-gap scenario, this entails \citep{rs75}
\begin{equation} \label{eq:somev}
\left( \frac {e \Delta V_{\text{max}}} {m_{e} c^2} \right)^{3} \frac {\hbar H} {2 m_{e} c R_{c}^2} \frac{ B_{p}} {B_{\text{QED}}} \gtrsim \frac {1}{15},
\end{equation}
where $B_{\text{QED}} = m_{e}^2 c^3/ e \hbar \approx 4.4 \times 10^{13}\,\mbox{G}$ is the Schwinger field for electron mass and charge $m_{e}$ and $e$, respectively, $\hbar$ is the reduced Planck constant, and $H$ denotes the gap thickness.  Note that the numerical factor $1/15$ in \eqref{eq:somev} depends weakly on the angle made between the direction of photon propagation and $\boldsymbol{B}$; see the discussion around equation (19) in \cite{rs75} for more details.  Moreover, the polar field strength $B_{p}$ may exceed the \emph{dipole} value, $B_{d}$, which is the relevant quantity at large radii. The curvature radius of a magnetic field line, $R_{c}$, scales inversely with the multipole order. Depending on the assumptions one places on the magnetic configuration, most notably $H$, $R_{c}$, and $B_{p}/B_{d}$, a variety of possible `death lines' arise.

There are two main radii characteristic to the magnetosphere of an isolated object, being the stellar radius and the light-cylinder radius, $\rLC = c P / 2 \pi$. \cite{cr93} posit that depending on the magnetospheric twist, $H$ and $R_{c}$ can assume a variety of values that are built from these two radii, such as $( \Rs \rLC )^{1/2}$, $ \Rs ( \Rs / \rLC)^{1/2}$, and so on. For polar-gaps, the thickness $H$ may also scale with the dimensionless ratio $\beta = B_{p}/B_{d}$, as multipoles can dominate over the dipole component near the stellar surface. In reality, a force-free magnetospheric model, though possibly with some displacement currents, is necessary to determine the gap size and curvature radius self-consistently for some field geometry. Nevertheless, we can explore the various extrema by taking the approximate scalings considered by \cite{cr93} and others. 

Following \cite{cr93} \cite[though see also][]{ha01,sz15}, there are four types of polar-gap scenario that we consider:
\begin{itemize}
\item[(a)]{Pure central dipole. This is the simplest such model, where $B_{p} = B_{d}$, $R_{c} = (\Rs \rLC)^{1/2}$ and $H = \Rs (\Rs / \rLC)^{1/2}$. The required field strength for pair-production is $B_{d,\text{min}}=2.2 \times 10^{12} ( P^{15/8} R_{6}^{-19/8})\,\mbox{G}$.}

\item[(b)]{Twisted dipole. As above, though instead $R_{c} = \Rs$. We find $B_{d,\text{min}}=2.7 \times 10^{11} (P^{13/8} R_{6}^{-17/8})\,\mbox{G}$.} 

\item[(c)]{Starspot configuration. Numerical simulations of crustal Hall drift tend to find that concentrated `magnetic spots' emerge near the polar-cap \cite[e.g.,][]{vig13,suv16}, where $B_{p} \gg B_{d}$. Taking $R_{c} = \Rs$ and a reduced cap size $H = \beta^{-1/2} \Rs ( \Rs / \rLC)^{1/2}$ yields $B_{d,\text{min}}=2.7 \times 10^{11} (\beta^{-1/8} P^{13/8} R_{6}^{-17/8})\,\mbox{G}$.} 

\item[(d)]{Twisted multipoles. Similar to case (c), though with maximum pitch angle between the magnetic field and the direction of emitted photons, $\sin \theta \approx H / R_{c}$ (and thus $H \approx R_{c} = \Rs$). Effectively, one assumes the magnetosphere is so twisted that a curvature-radiation photon, emitted almost parallel to $\boldsymbol{B}$, crosses another part of the cap's open field line bundle at a large angle. This gives $B_{d,\text{min}}=9.2 \times 10^{10} (\beta^{-1/4} P^{3/2} R_{6}^{-2})\,\mbox{G}$.} 
\end{itemize}

\begin{table}
	\centering
    \caption{Minimum polar dipole field strengths required for GLEAM-X J1627, assuming a death line according to the characterisation given in the main text. The final column shows $B_{d,\text{min}}$ for a canonical radius $R_{6} = 1$, with the bracketed number corresponding to a larger radius of $13\, \mbox{km}$. The second column indicates the strength of the surface field relative to the dipole component, which influences the cap thickness, $H$, and curvature radius, $R_{c}$.}
    \label{tab:deathlines} 
    \begin{tabular}{lcc}
    \hline
    \hline
     Magnetic geometry & $B_{p}/B_{d}$ & $B_{d,16}^{\text{min}}$ $(R_{6} = 1.3)$\\
      \hline
    (a) Pure dipole & 1 & 101 (58.8)\\
      \hline
    (b) Twisted dipole & 1 & 2.30 (1.32)\\
      \hline
    (c) Starspot & 2 & 2.10 (1.21)\\
   &  5 & 1.88 (1.08)\\
    & 10 & 1.72 (0.99)\\
      \hline
     (d) Twisted multipoles & 2 & 0.279 (0.165)\\
     & 5 & 0.222 (0.131)\\
     & 10 & 0.187 (0.111)\\
      \hline
    \end{tabular}

\end{table}

Table \ref{tab:deathlines} displays the minimum polar dipole strength $B_{d,\text{min}}$, in the context of the death lines (a)--(d) described above, for a range of surface-to-dipole ratios, $B_{p}/B_{d}$, and a canonical radius $\Rs = 10\, \mbox{km}$, such that GLEAM-X J1627 can activate as a radio pulsar. Taking instead a stellar radius of $13\, \mbox{km}$ allows for an easier switch on, as shown by the numbers in parentheses. Rows show different values for the relative strength of multipoles, which influence the curvature radius and gap thickness, as described above. Although lines (b) through (d) imply high multipole orders ($\ell \gtrsim 10^2$) when interpreted via the numerical simulations of \cite{ass01}, {for instance}, magnetohydrodynamic evolutions in proto-magnetars indicate that the generation of high-order multipoles in the interior is a generic quality of strongly-magnetised systems \citep{kiu18,hask21}. \cite{hask21} found that truncating their numerical output to harmonic expansions with $\ell_{\text{max}} < 32$ led to sizeable inaccuracies in the inferred field strength (see also Sec. \ref{sec:hall}). Fallback accretion onto the proto-star, or later in life, can also twist field lines near the stellar surface \citep{mel14,suvm20}.

{We emphasise that the above models, while phenomenological, represent the extrema that could be expected for a given activation mechanism. It is unlikely that any given line applies to the entirety of the neutron star population at all times. For example, magnetospheric twists (lines b and d) and starspots (line c) are dynamical in nature and subject to diffusion, implying that the radius of curvature is a function of time. Regarding magnetars in particular, \cite{belo09} suggests that their radio activation may be related to quake activity in the crust, where overstressed zones fracture and flow plastically, dragging the magnetic footpoints with them and pumping a current (`$\boldsymbol{j}$-bundle') into the magnetosphere \cite[see also][]{bt07}. Overshearing events may also be expected from spindown \citep{baym71}, which tends to be faster in magnetars. Magnetospheric twists can survive on $\gtrsim$ year-long timescales \citep{parf13}, until the field lines relax to the pre-twist (or some other) equilibrium. During a twisting episode one may expect that either lines (b) or (d) apply, after which a dipole or starspot configuration is reinstated (lines a or c) depending on the surface-field multipolarity. This may help to explain why certain magnetars are radio loud while others are not, depending on the waiting time between twist injections \cite[see also][]{mor12}.} 

\begin{figure*}
\includegraphics[width=2\columnwidth]{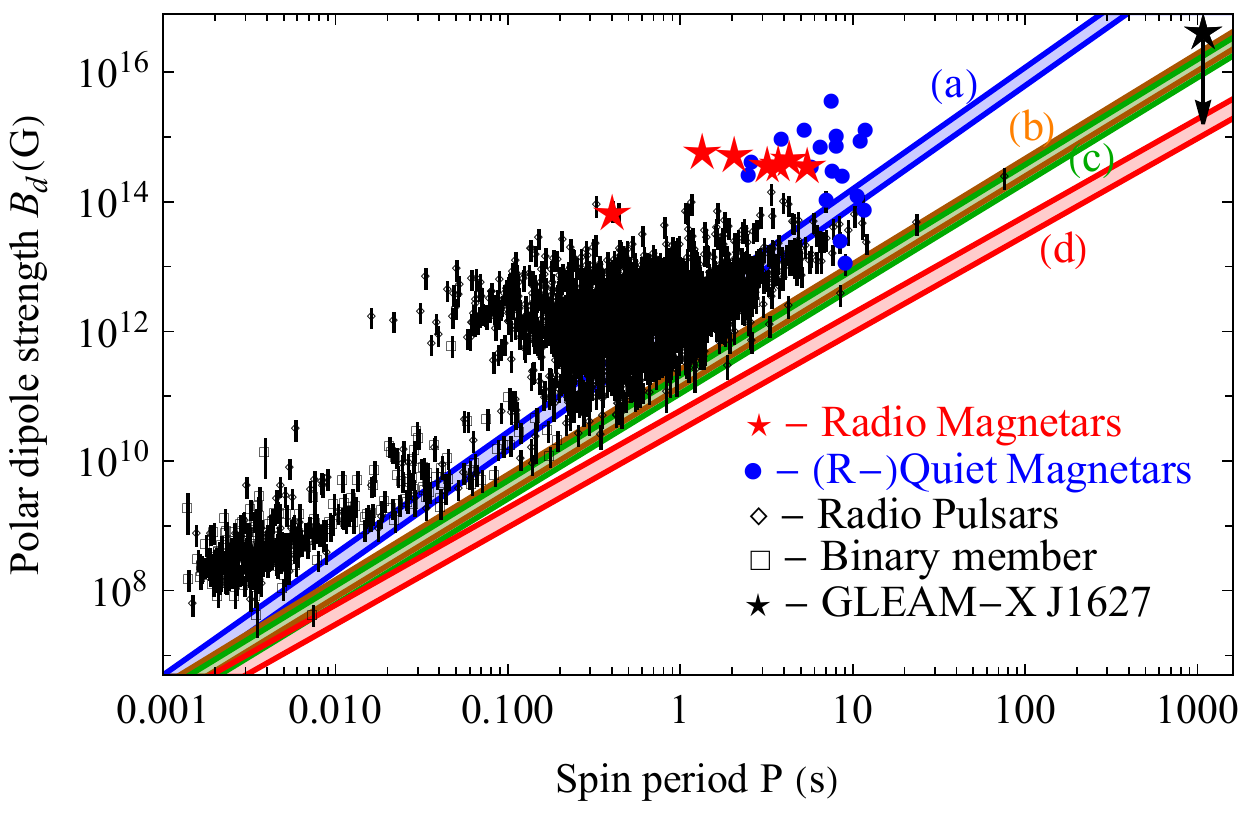}
\caption{$B_{d}-P$ diagram overlaid with death `lines' (a)--(d), as shown by the coloured curves (see plot legends). Each curve comes with some thickness because we allow for uncertainties in the stellar radius, $10 \leq R_{\star}/{\rm km} \leq 13$, and the polar-to-dipole field strength ratio, $1 \leq \beta \leq 10$. Overlaid are known objects, with available $\dot{P}$ measurements, from the ATNF pulsar catalogue \protect\citep[\protect\url{http://www.atnf.csiro.au/research/pulsar/psrcat};][]{atnf} with `ordinary' radio pulsars shown by black diamonds, those in binaries with black squares (for which $B$-field estimates are even more uncertain as they depend on accretion assumptions), radio-loud magnetars with red stars, and radio-quiet magnetars with blue circles. {The (vacuum dipole) upper limit for GLEAM-X J1627 is indicated with a black star}. The $B_{d}$ values for other pulsars come from the standard dipole-braking formula \eqref{eq:braking} with $n=3$ for a range of obliquities ($\pi /4 \leq \alpha \leq \pi/2$); these variations, together with those on $R_{\star}$ and timing errors on $\dot{P}$, imply some uncertainty on the dipole strength. For magnetars {(except J1119)}, mean values from the McGill catalogue are used \protect\citep{mcg14}. In principle, all objects lie above the overall valley, defined as the area between the top of line (a) and the bottom of line (d), with the possible exception of GLEAM-X J1627, unless it has a highly-twisted magnetosphere ($R_{c} \sim R_{\star}$) with a polar \emph{dipole} field strength exceeding $\sim 10^{15}$G.}
\label{fig:death}
\end{figure*}

{Detailed simulations of pair cascades in magnetar environments were carried out by \cite{med10} to examine whether the polar gap story, discussed above, applies in super-strong fields \cite[see also][]{td96,bar99,bt07}. Although they find the cascade proceeds differently for fields above and below $B_{\rm QED}$, mostly because of the suppression of synchrotron emission in strong fields, the overall multiplicity $\lambda$ is relatively insensitive to $B$. The energy spectrum of electron-initiated cascades depends mostly on the polar-cap voltage, and hence the spin period, and not $B$ alone. The critical multiplicity, necessary for radio activation, that they find in the strong-field case is $\lambda_{\rm death} \approx 1.5 \times 10^{7} (B_{p}/10^{12}\text{G})^{-1/6} (R_{c}/10^{8}\text{cm})^{2/3}$ (see their Sec. 4.2). For untwisted dipoles this implies $B_{d,\text{min}} \propto P^2$, similar to line (a) though marginally steeper. For highly-twisted fields with $R_{c} \sim R$, one recovers lines qualitatively similar to either (b) or (d) from their results, depending on the field topology. Therefore, although the overall slope of death lines in $B$-$P$ space may vary depending on how the cascade proceeds, the death valley defined as the area spanned by lines (a) through (d) is a reasonable approximation for the valley extrema, even for magnetars\footnote{{Given the high degree of nulling and that only bright and variable single pulses were detected from GLEAM-X J1627, it may be that the radio activity is not attributable to traditional cascades. Different death lines altogether may apply, such as the fast radio burst death lines described by \cite{wad19}, which can be satisfied with somewhat weaker fields (see their equation 9).}}.}

{Figure \ref{fig:death} shows death lines in the context of the wider neutron star population. Although noting the caveats discussed above, we see that all known, pulsating objects lie above the overall valley, with the possible exception of GLEAM-X J1627 (shown by a black star). The vertical axis shows the dipolar field strength of known objects, which is relatively uncertain: one requires a braking model to estimate this quantity. We assume that the standard magneto-dipole picture [equation \eqref{eq:braking} with $n=3$] applies, though allow for uncertainties in the inclination angle, $\pi/4 \leq \alpha \leq \pi/2$, and the stellar radius, $10 \leq \Rs/{\text{km}} \leq 13$. Note that  polarisation data from radio pulsars indicate that $\alpha$ spans an even larger range \cite[e.g.,][]{rank90}. Therefore, individual death lines have some width, and pulsar positions on the diagram have some uncertainty (see Tab. \ref{tab:deathlines}). Although the mean of line (a) cuts right through the middle of the population, inclinations tending towards alignment weaken the intrinsic torque, and thus predict a larger $B$ for a given $\dot{P}$. It was argued by \cite{cont06} that radio pulsars may evolve towards an aligned configuration $(\alpha \rightarrow 0)$, and hence even line (a) could be sufficient in most cases. Evolution towards alignment is also observed in 3D magnetospheric simulations \citep{phil14} \cite[though cf.][]{land18}. For many systems however it is likely that dynamical phenomena, such as magnetospheric twist injections via magnetically- \citep{belo09} or spindown-induced \citep{baym71} quakes, or starspot formation \citep{zhang07}, play a role in pair cascade phenomena. Lines (b) through (d) may therefore only apply sporadically over $\sim$ year-long timescales \citep{parf13}.}

{In the context of Fig. \ref{fig:death}, we see that the} pure dipole [model (a)] is unable to explain the {radio switch-on} of GLEAM-X J1627 unless the polar field takes on super-virial values, $B_{p} \gtrsim 10^{18}\,\mbox{G}$. This is in contrast with all (other known) radio-loud magnetars, namely {PSR J1745--2900, PSR J1622--4950, XTE J1810--197, 1E 1547.0--5408, Swift J1818.0-1607, {PSR J1119--6127} and SGR 1935+2154 (red stars)}. {This line fails to explain the pulsar population at large though, cutting through the middle of the diagram.} Only in the case of a highly-twisted configuration [model (d)] can the local field of GLEAM-X J1627 assume values $B_{p} \lesssim 10^{16}\,\mbox{G}$. For these models, the minimum required for the surface field is still greater than the maximum polar field amongst all other 31 known magnetars, with the runner up being SGR 1806--20 \citep{mcg14}, an extraordinarily bright and young ($<\, \mbox{kyr}$) burster, which boasts a polar field strength $B_{p} \approx 4 \times 10^{15}\,\mbox{G}$. 

The uncertainty implied by the final column of Tab. \ref{tab:deathlines} is a lower limit, as consideration of outer gap models \citep{cr93}, thermionic emissions \citep{sz15}, and general-relativistic corrections \citep{ha01} can also adjust the voltage drop. Outer-gap models, however, tend to fare worse. For example, for the partially-inclined, outer magnetosphere accelerator model [Eq. (27) of \cite{cr93}], the relevant death line, $5 \log B_{p} - 12 \log P \approx 69.5$, demands a magnetic field for GLEAM-X J1627 that exceeds the virial limit. \cite{sz15} argued, in the context of a partially-screened gap, that the polar cap must be below some critical $B$-dependent temperature, else thermionic emissions effectively screen the acceleration potential (such considerations are pertinent to the observational absence of X-ray emissions; see Sec. \ref{sec:xrays}). Similarly, the general-relativistic Lense-Thirring corrections discussed by \cite{ha01} may be important for GLEAM-X J1627, despite its long spin period, because the \cite{gj69} plasma density and the Lense-Thirring frequency both scale linearly with the rotation rate. Na\"{\i}vely applying the `low-altitude' estimate for the pair multiplicity computed by \cite{ha01}, which includes Lense-Thirring precession and is appropriate when $B_{p}/(10^{12}\, \mbox{G}) \gtrsim P/(1\, \mbox{s})$ [see their Eq. (69)], we obtain a minimum polar field strength $B_{p,\text{min}} \approx 2 \times 10^{16}\, \mbox{G}$, comparable to lines (b) and (c). The \cite{ha01} models however invoke cap temperatures set by backflowing positrons, which may be unrealistically low for GLEAM-X J1627 because internal heating driven by field decay is likely to be non-negligible (see Sec. \ref{sec:xrays}); thermal transport simulations would be necessary to self-consistently assess the valley structure in this case.

\subsection{Hall-plastic-Ohm decay}
\label{sec:hall}

A simplified picture of the neutron star crust is that of a rigid, ion lattice strewn with mobile electrons. The latter carry a current as they flow relative to the ions, gradually advecting the field lines that thread the crust. This process of Hall drift, while conserving magnetic energy, can act to accelerate Ohmic decay through a sequence of cascades to smaller-scale magnetic structures, possibly aided further by thermoelectric effects \cite[see][for a review]{g22}. The Hall timescale obeys $\tH \propto B^{-1}$, and thus magnetar crusts are particularly prone to field decay. {Depending on the initial conditions however, the system may enter into an `attractor' state where the Hall term vanishes \citep{gour14}. Although we will not consider this complication further, a Hall-stalled evolution may help GLEAM-X J1627 to maintain a strong field while cooling quickly as it ages.} 

As magnetic gradients form, Maxwell stresses are exerted on the crust. For magnetar-like field strengths $B \gtrsim 10^{15}\,\mbox{G}$, the crust may not be sufficiently malleable to absorb these stresses, and rather a crustal failure may occur \citep{duncan98,lan15}. Crustquakes are popular models for the progenitors of magnetar outbursts, such as giant flares \cite[e.g.,][]{gog00} or fast radio bursts \cite[e.g.,][]{suvk19}. Once the crust experiences a failure however, it is unlikely to `heal' immediately and rather may enter a state of azimuthal shearing termed plastic flow \citep{bell14,lg19,koj22}. Plastic flow is generally a dissipative process, and thus depending on the `plastic viscosity', the Hall effect may be enhanced, implying that numerical Hall-Ohm (as opposed to Hall-plastic-Ohm) investigations underestimate the degree of field decay. On the other hand, plastic flows can move against the existing flow of the electron fluid \citep{gl21}, and thus inhibit magnetic dissipation by counteracting the formation of the small-scale (i.e., highly multipolar) magnetic substructures most susceptible to Ohmic decay. The density-dependent, and hence radially-stratified, nature of the electron fluid flow also facilitates the growth of a toroidal field, making an investigation of a realistic Hall-plastic-Ohm system a challenging task.

The evolution of the crustal magnetic field $\bb$ is described by the induction equation
\begin{equation} \label{eq:induction}
\begin{aligned}
0 =&\frac {\partial \bb} {\partial t} + \nabla \times \Big[ \frac {c} {4 \pi e n_{e}} \left( \nabla \times \bb \right) \times \bb  \\
&- \bvPL \times \bb + \frac {c^2}{4 \pi \sigma} \nabla \times \bb \Big],
\end{aligned}
\end{equation}
for electron number density $10^{34} \lesssim n_{e}/ \text{cm}^{-3} \lesssim 10^{36}$ and conductivity $10^{16} \lesssim \sigma / \text{s}^{-1} \lesssim 10^{24}$, where $\bvPL$ denotes the plastic flow velocity. The lower limit for the electrical conductivity applies to the crust-magnetosphere interface, while the latter is appropriate for the inner crust \citep{akg18}. A proper description for $\bvPL$, including a determination of characteristic plastic speeds, requires an additional equation of motion, typically set by the requirement that a Stokes flow is induced in regions of excess stress \cite[determined, e.g., by the von Mises criterion;][]{lan15}. Following \cite{ag08}, we construct an approximate model by replacing the gradient operator with the inverse of a relevant lengthscale, $L$, yielding {\cite[see also][]{lan22}}
\begin{equation} \label{eq:hallohm}
\frac {d B} {dt} = - \frac {B} {B_{0}} \frac{B} {\tH} + \frac {\vPL B} {L} - \frac {B} {\tO},
\end{equation}
with $\tH = 4 \pi e n_{e} L^2 / c B_{0}$ and $\tO = 4 \pi \sigma L^2 /c^2$ read off from \eqref{eq:induction}, with small-scale structures dominating the choice of $L$. Equation \eqref{eq:hallohm}, which is subject to the initial condition $B(0) = B_{0}$, reduces to the phenomenological Hall-Ohm model of \cite{ag08} when $\vPL = 0$.

In the simulations of \cite{lg19}, it was found that larger $B_{0}$ values lead to swifter plastic flows, and more precisely that doubling $B_{0}$ leads to an approximately 3-fold increase in $\vPL$. For magnetar-level fields and low plastic viscosities, these authors \cite[see also][]{gl21,g22} found that $\vPL$ can approach a few hundred cm per year in regions where field lines are particularly tangled. However, slower plastic speeds emerge in the bulk of the crust, and no flow at all occurs in unstressed regions. As we have washed out all spatial dependencies in building relation \eqref{eq:hallohm}, the flow is nominally non-zero everywhere, rather than only in regions localised around failures. We thus consider instead spatially-averaged speeds of $\vPL \lesssim 40 \text{ cm yr}^{-1}$ for cases with ultra-strong fields. Note that if $\vPL$ is negative (i.e., if one takes $\nabla \rightarrow -1/L$ rather than $\nabla \rightarrow 1/L$), plastic flow instead accelerates field decay; such cases have been observed in the studies cited above, depending on the plastic viscosity.

\begin{figure}
\includegraphics[width=\columnwidth]{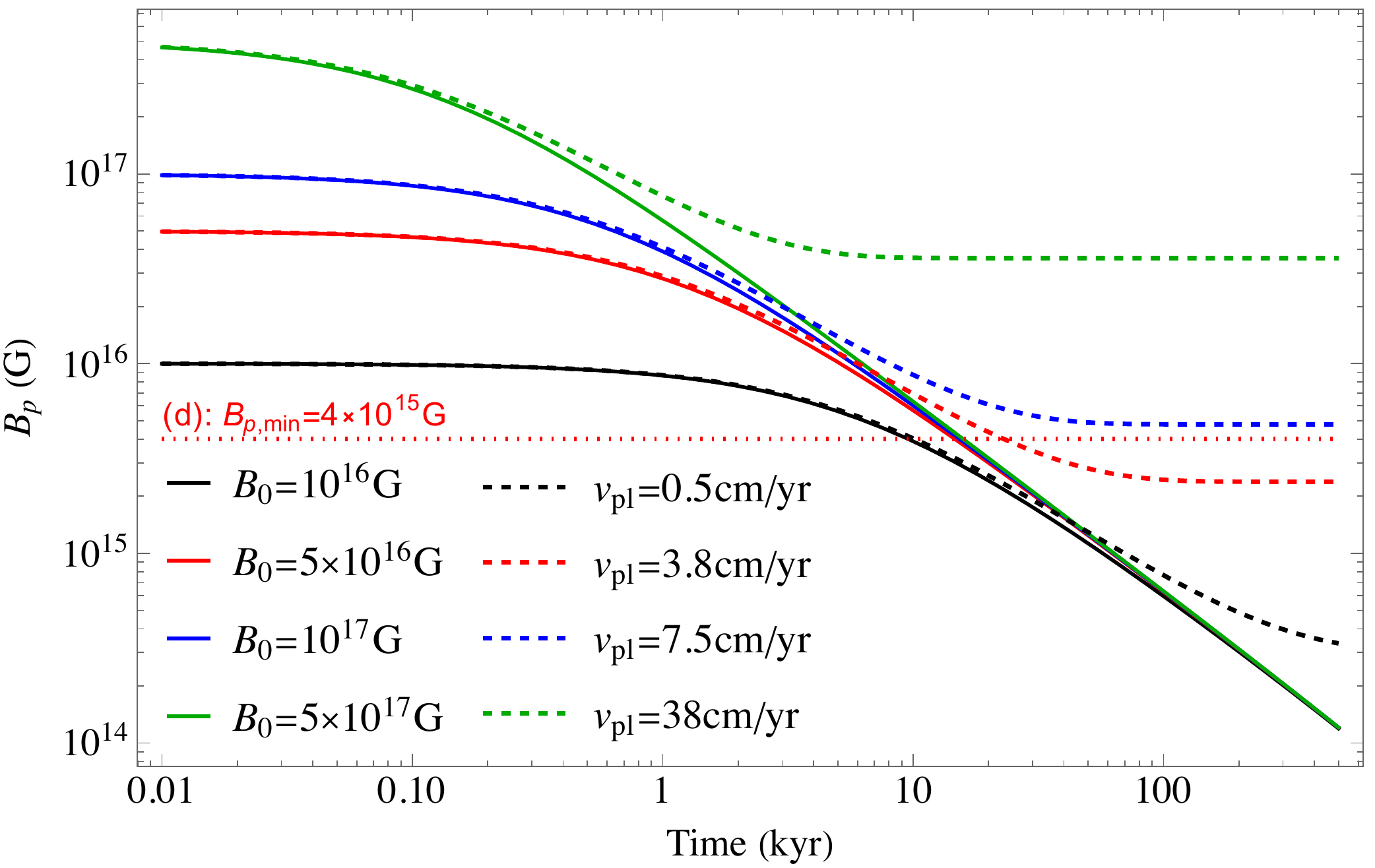}
\caption{Evolution of the polar field strength, $B_{p}(t)$, given as a solution to equation \eqref{eq:hallohm} for both Hall-Ohm ($\vPL = 0$, solid curves) and Hall-plastic-Ohm ($\vPL \neq 0$, dashed curves) evolutions, for several birth field strengths and plastic velocities (see colour-coded legends). The dotted, horizontal line shows the minimum field strength set by the type (d) death lines with $B_{p}/B_{d} \sim 2$ considered in Sec. \ref{sec:deathval}.}
\label{fig:hallohm}
\end{figure}

Figure \ref{fig:hallohm} shows solutions to equation \eqref{eq:hallohm} for several initial field strengths $1 \leq B_{16} \leq 50$, in both the Hall-Ohm \cite[solid curves]{ag08} and Hall-plastic-Ohm (dashed curves) cases, where the plastic flow velocity is chosen to scale with $B$ in the manner described above. To provide an optimistic but realistic\footnote{Note that, because $|\nabla B| / B \propto \ell^{-1}$ for a general $\ell$-pole, if one were to assume a purely dipole field for all $t$, a longer lengthscale $L \lesssim \Rs$ could be justified, which would extend the Hall time. Such an assumption would, however, be inconsistent with the twisted surface configurations studied for death valleys in Sec. \ref{sec:deathval}.} scenario, we set $L = 10^{5}$~cm, $n_{e} = 10^{36}$, and $\sigma = 10^{24} \text{ s}^{-1}$; smaller values lead to faster decays. When $\vPL \leq 0$, the field enters a state of rapid decay after $\sim 1$~kyr, reducing by an order of magnitude after only $\approx 2$~kyr in the ultra-strong case with $B_{0} = 5 \times 10^{17}$ G. The dotted line illustrates the minimum field strength required to fulfil the death valley requirements discussed in Sec. \ref{sec:deathval}. We remind the reader that even if the \emph{dipole} field is of order $B_{d} \sim 10^{15}\,\mbox{G}$, {the surface field implied by the death valley constraints is of order $\gtrsim 4 \times 10^{15}\,\mbox{G}$; see Tab. \ref{tab:deathlines}.}

Demanding that $B_{p} \gtrsim 4 \times 10^{15}\,\mbox{G}$ at present implies that the system can be at most $\sim 20\, \mbox{kyr}$ old independently of the birth field strength if plastic flow is ignored, because one has $\tH \propto B_{0}^{-1}$. Such a conclusion may be in tension with the observed spin period of GLEAM-X J1627 {(see Sec. \ref{sec:brake})}, unless the star underwent a period of extreme spin-down early in its life \cite[from, e.g., propellering fallback material shortly after birth, as discussed by][]{hw22b,genc22}. Including a sufficiently rapid plastic flow, $\vPL \gtrsim 30 \text{ cm yr}^{-1}$, however stalls the impact of the Hall effect \citep{gl21}, allowing the field to decay only on the true Ohmic timescale, $\tO \gg 10^{2}\, \mbox{kyr}$. In this case, field strengths of order $\gtrsim 5 \times 10^{15}\, \mbox{G}$ can be maintained over relatively long timescales if $B_{0} \sim 10^{17}\,\mbox{G}$. 

\subsection{Braking mechanism}
\label{sec:brake}

Neutron stars in isolation spin down gradually as electromagnetic and gravitational torques are applied, the magnitude of which can be phenomenologically quantified in terms of a \emph{braking index}, $n$. For a centered dipole {that never decays} one has $n=3$, while for a general $\ell$-pole we have $n = 2 \ell + 1$. Leading-order contributions from gravitational radiation, being quadrupolar, also give $n = 5$, though with a different prefactor. Values $n<3$ are also possible for an oblique and/or precessing rotator \citep{mel97,mel99}, or in cases where particle outflows dominate the spin-down torque \citep{cont99,thomp00}. It is therefore useful to consider an evolution with an arbitrary braking index. 

The spin evolution of an inclined rotator in vacuum can be described by \cite[e.g.,][]{man77}
\begin{equation} \label{eq:braking}
\dot{P} = \left(2 \pi \right)^{n-1} \frac {B_{p}^2 \sin^2 \alpha P^{2-n} \Rs^{3+n}} {6 I_{0} c^n},
\end{equation}
for moment of inertia $I_{0}$ and  magnetic inclination angle $\alpha$. {The quantity $n$ in \eqref{eq:braking} represents the observational braking index only if $B_{p}$ is constant, though in the absence of $\ddot{P}$ data we treat it as phenomenological.} We adopt the general-relativistic Tolman VII equation of state, for which $I_{0} = 0.38 M_{\star} \Rs^2$ \citep{lp01} for stellar mass $M_{\star}$. Equation \eqref{eq:braking} provides two useful pieces of information. Firstly, the present-day observations of $P$ and $\dot{P}$ provide an estimate for $B_{p}$ for a given braking index. Secondly, by solving equation \eqref{eq:braking} for some (time-dependent) choices of $n$ and $B_{p}$, one can infer the age of the system. Given that we anticipate the object was born rapidly rotating so as to explain its large field strength (see Sec. \ref{sec:birth}), its present-day period must far exceed its birth period $P_{0}$, though age ($\tau$) estimates from equation \eqref{eq:braking} are insensitive to $P(0)$ for $P(0) \ll P(\tau)$. Note that magnetospheric \citep{spit06,phil14}, spheroidal, {general relativistic}, or offset corrections \citep{pet22} can be accounted for in the above to adjust the effective $B_{p}$ value; one obtains a hybrid \cite{spit06} formula, for example, by replacing $B_{p}^2 \sin^2 \alpha$ in expression \eqref{eq:braking} with $B_{p}^2 (1 + \sin^2 \alpha)$.

{We solve equation \eqref{eq:braking} simultaneously with the volume-averaged induction equation \eqref{eq:hallohm} for several values of $\dot{P}(\tau) \leq 1.2 \times 10^{-9} \text{ss}^{-1}$ \citep{hw22} assuming an orthogonal rotator, $\alpha = \pi/2$. We fix $n$ by demanding $B_{p}(\tau) = 5 \times 10^{15}\, \mbox{G}$, as it is difficult to explain the present-day radio switch on if the field is weaker (see Fig. \ref{fig:death}, keeping in mind the caveats noted in Sec. \ref{sec:deathval}).  The three \emph{a priori} free parameters, namely $B_{0}$, $\tau$, and $n$, are uniquely determined by the specified values of $\dot{P}(\tau)$, $B_{p}(\tau)$, and $P(\tau)$. Solutions are built through a shooting method: a set of initial conditions are iteratively determined such that there exists an age $\tau$ for which the aforementioned conditions are met.} 

\begin{figure}
\includegraphics[width=\columnwidth]{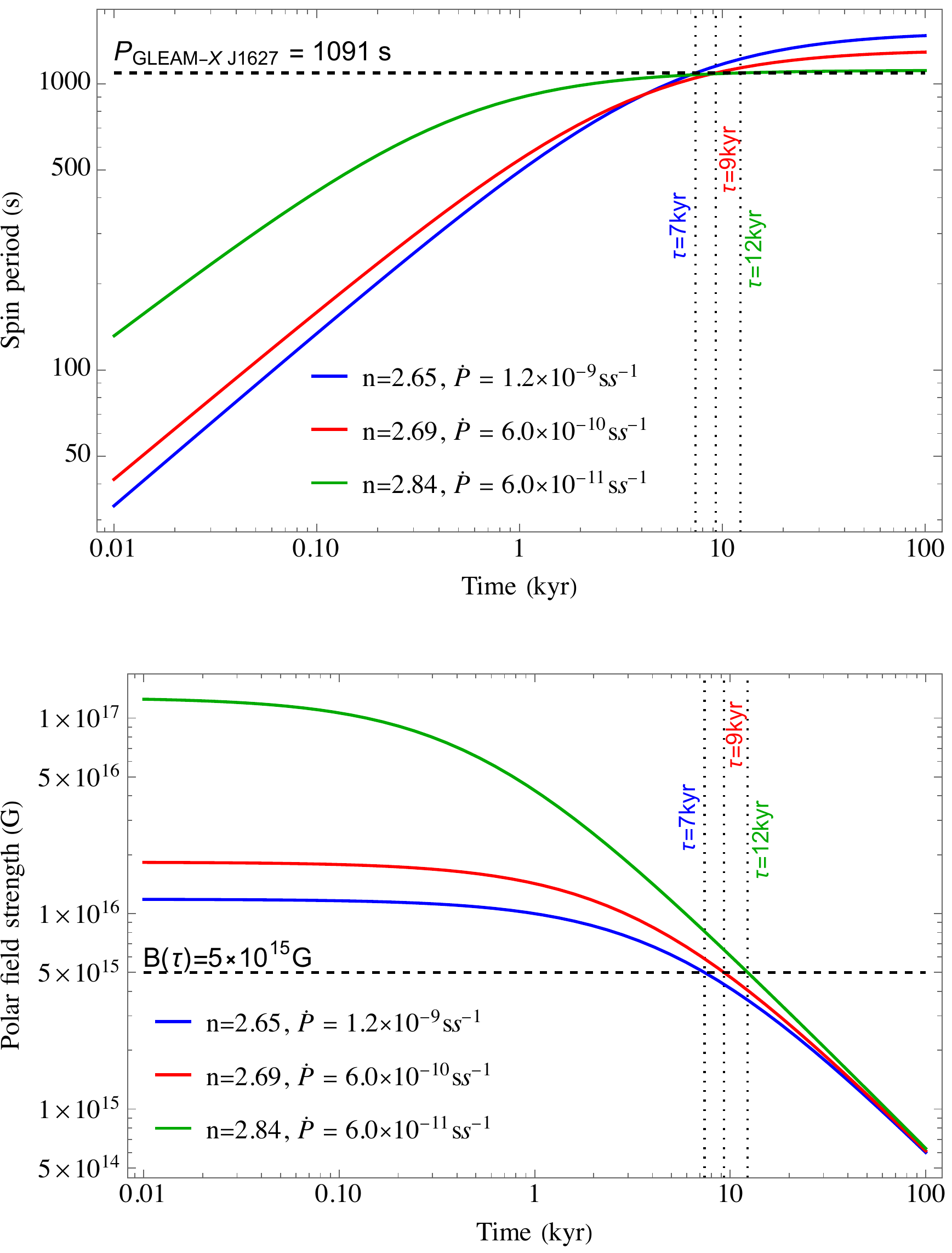}
\caption{Solutions to equation \eqref{eq:braking} for the rotational period $P(t)$ (top panel), assuming a time-dependent magnetic field $B_{p}(t)$ whose evolution is governed by \eqref{eq:hallohm} (bottom panel), for a variety of $\dot{P}(\tau)$ values (see plot legends). The plastic velocity is set to zero in these examples. The age, braking index, and birth field strengths are set by the conditions that $P(\tau) = 1091.17$s and $B_{p}(\tau) = 5 \times 10^{15}$G for some given value of $\dot{P}(\tau)$.}
\label{fig:brake1}
\end{figure}

{Figure \ref{fig:brake1} shows the evolutions of the spin period (top panel) and the polar magnetic field (bottom) for cases where plastic flow is ignored. Even for a relatively large range of the present-day period derivative, $6.0 \times 10^{-11} \leq \dot{P}(\tau)/(\text{ss}^{-1}) \leq 1.2 \times 10^{-9}$, the evolutions proceed in a similar manner. This occurs because we require that the present-day $B$ field is still strong, $B_{p}(\tau) = 5 \times 10^{15}$G, so as to accommodate the death valley minima discussed in Sec. \ref{sec:deathval}. In the run with $\dot{P}(\tau) = 6.0 \times 10^{-11} \text{ ss}^{-1}$, for example, the birth field strength must exceed $10^{17}$G so that it can survive long enough (until $\tau = 12$kyr) to ensure that the present-day switch-on minimum is met, which implies greater spindown during early times $t \ll \tau$. As such, even if a factor $\gtrsim 10$ weaker $\dot{P}(\tau)$ is assumed than the best-fit value reported by \cite{hw22}, predictions for the age are quantitatively similar in cases where plastic flow and other torques are inactive \cite[cf.][]{hw22b,genc22}.
}

\begin{figure}
\includegraphics[width=\columnwidth]{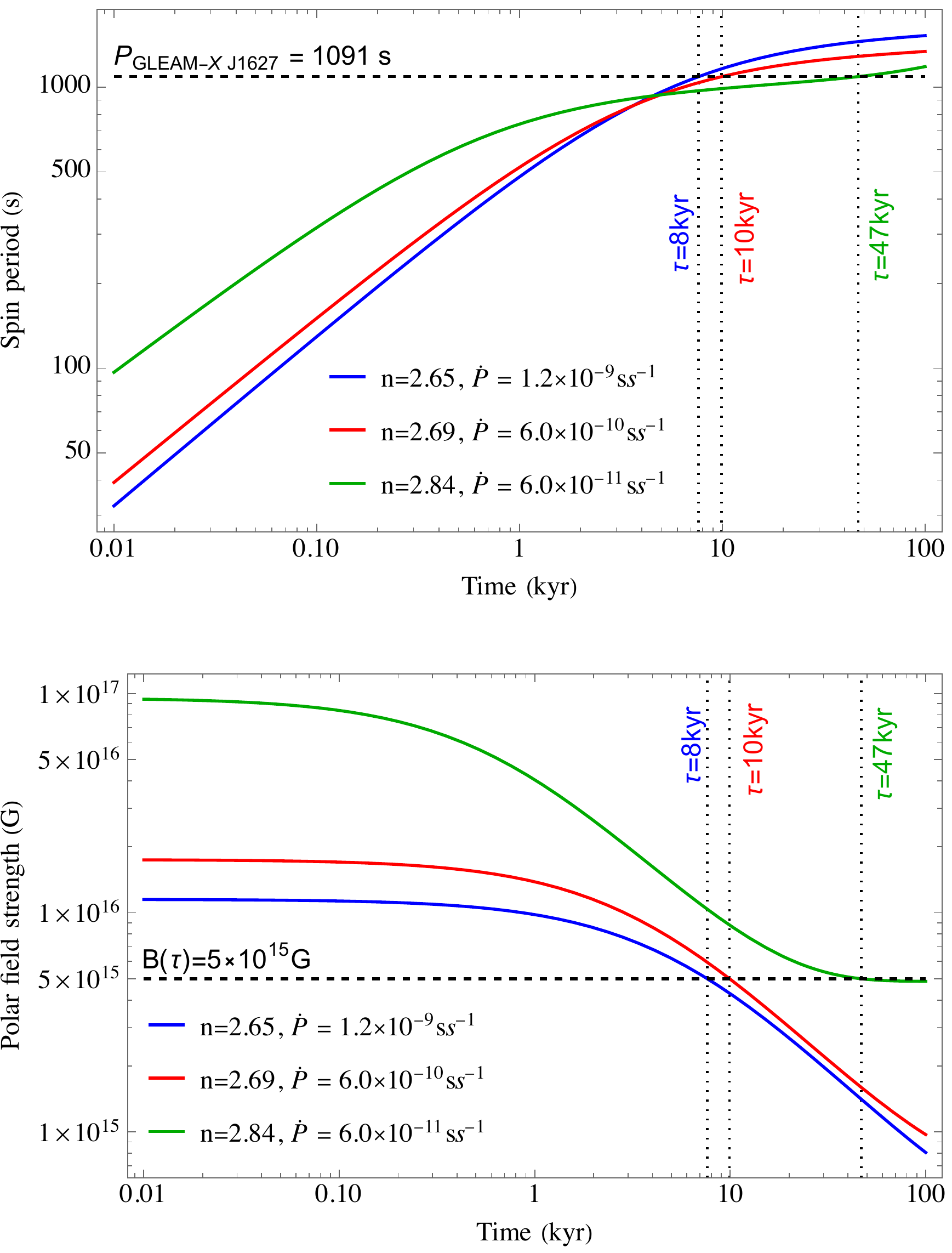}
\caption{Similar to Fig. \ref{fig:brake1}, though with a non-zero plastic velocity whose value is set by the scaling discussed in Sec. \ref{sec:hall}, i.e., doubling $B$ relative to some fixed value implies a 3-fold increase in $v_{\rm pl}$, where we set $v_{\rm pl} = 0.5 \text{ cm yr}^{-1}$ for $B_{0} = 10^{16}$G. }
\label{fig:brake2}
\end{figure}

{By contrast, evolutions carried out for $v_{\rm pl} \neq 0$ are shown in Fig. \ref{fig:brake2}. In these cases, $\dot{P}(\tau)$ makes a significant difference for the age prediction: for $\dot{P}(\tau) = 1.2 \times 10^{-9} \text{ ss}^{-1}$ we find $\tau = 8$kyr, while for the smaller value $\dot{P}(\tau) = 6.0 \times 10^{-11} \text{ ss}^{-1}$ the age prediction increases to $\tau = 47$kyr. This is because plastic flow stalls field decay (see Fig. \ref{fig:hallohm}), allowing the star to match $B_{p}(\tau) = 5 \times 10^{15}$G without having to be born with a field exceeding $10^{17}$G. In this way, spindown is slower in the early stages and the star can be older. Increasing the plastic velocity can increase the age further; in the limit that $v_{\rm pl} \rightarrow \infty$ (or $\tau_{\rm Ohm} \rightarrow \infty$) the field does not decay at all, and the age is simply given by the characteristic value $\tau \propto P(\tau)/ \dot{P}(\tau)$, which can be arbitrarily large if $\dot{P}$ tends towards zero.}

\subsection{Birth conditions: field amplification}
\label{sec:birth}

{To a large degree, it remains an open question as to how magnetars acquire their intense fields, especially large-scale dipoles. The saturation amplitude of the core field in the case of dynamo activity shortly after birth could reach $\lesssim 10^{16}\,\mbox{G}$ for convective heat fluxes of order $\gtrsim 10^{39} \text{ erg cm}^{-2} \text{s}^{-1}$ \citep{td93}. Provided that an `inverse cascade' can operate, where energy from turbulent patches is transferred into a large-scale dipole \cite[cf.][]{jank17,ray20}, birth fields of this magnitude are sufficient for all of the known Galactic magnetars.} 

Mechanisms beyond dynamo activity can amplify a magnetic field. In particular, the Kelvin-Helmholtz and magneto-rotational instabilities can potentially lead to saturation magnetic energies of order $U_{\text{mag}} \sim 10^{51}$ erg provided that the star is born with a (sub-)millisecond period \citep{kiu18,c20b,ciolfi20,shib21}. We stress however that numerical studies reporting such large magnetic energies do so in the context of \emph{merger remnants}, which generally possess more angular momentum and seed magnetic fluxes than stars born from core-collapse. Regardless, magnetic energies of this order imply an upper limit to the volume-averaged magnetic field strength at birth, viz.
\begin{equation} \label{eq:sat}
\langle B \rangle_{\text{max}} \approx 7.7 \times 10^{16} \left( \frac{ U_{\text{mag}}} {10^{51} \text{ erg}} \right)^{1/2} \left( \frac{\Rs}{10 \text{ km}} \right)^{-3/2} \text{ G}.
\end{equation}

If the core field is at least as strong as the surface one (see also Sec. \ref{sec:xrays}), expression \eqref{eq:sat} implies that $B_{p} \lesssim \langle B \rangle_{\text{max}}$. The numerical simulations referenced above therefore suggest it is difficult to justify values exceeding $B_{p}(t=0) \sim 10^{17}\,\mbox{G}$ \cite[though cf.][]{suvg22}, even if toroidal fields \citep{glamp15} or intense magnetic spots \citep{vig13,suv16} are localised in the crust. 

\section{Is the magnetar hypothesis excluded by the absence of X-rays?}
\label{sec:xrays}

{Follow-up searches were carried out with the Swift X-ray Telescope for 2\,ks. An upper limit of $F_{\rm X} < 1.9 \times 10^{-13}\, \mbox{erg s}^{-1} \mbox{cm}^{-2}$, is found for the flux in the 0.3--10keV band, with $F_{\rm X} < 1.5 \times 10^{-13}\, \mbox{erg s}^{-1} \mbox{cm}^{-2}$ applying instead for a blackbody fit\footnote{{Note that, in general, power-law components and not just one or more blackbodies are also needed to fit magnetar spectra, see Table 2 in \cite{coti18}. For 4U 0142+61, for example, blackbody emissions represent $\sim 25\%$ of the total X-ray power \citep{rea07}. Spectral complications can be accounted for crudely in the models here via the efficiency parameter $\epsilon$ introduced below.}} at $kT \sim 0.1$~keV. Based on the greatest distance allowed by the dispersion measure, $d_{\text{max}} = 1.8\, \mbox{kpc}$, this gives $L_{\rm X} \leq 7 \times 10^{31}\, \mbox{erg s}^{-1}$ for the X-ray luminosity \citep{hw22}.} {A followup search conducted by \cite{rea22} implies an even tighter upper-limit for this $d_{\rm max}$, $L_{\rm X} \leq 2 \times 10^{30}\, \mbox{erg s}^{-1}$}. In this section, we review models of thermal regulation in magnetars as a means to predict the surface temperature as a function of field strength (Sec. \ref{sec:heatcold}), which is quantitatively applied to GLEAM-X J1627 in Sec. \ref{sec:thermres}.

\subsection{Heating and cooling}
\label{sec:heatcold}

The absence of thermal X-rays in particular poses a challenge to the magnetar interpretation of GLEAM-X J1627: the magnetised electron-proton plasma in the core experiences friction with the approximately static neutron fluid, gradually heating up the star while depleting magnetic energy \citep{gr92}. 

Ambipolar heating, which sets a floor value to the temperature for a given age, is counteracted by neutrino cooling \citep{tur11,ho12,vig13,anz21,anz22}. There is, therefore, a quasi-static balance temperature, $T_{\text{bal}}$, set by matching the (time-dependent) heating and cooling rates, which generally must be several times $10^{8}\, \mbox{K}$ to explain observations from active magnetars \citep{td96,belo16}.

{Performing a volume average, the core temperature evolution can be approximately described by the first law of thermodynamics,}
\begin{equation} \label{eq:tmod}
C_{\rm V} \frac {d T_{\rm core}} {dt} = \dot{Q}_{B} - \dot{Q}_{\nu},
\end{equation}
{for heat capacity $C_{\rm V} \approx 2 \times 10^{20} (T_{\rm core}/10^9 \text{K}) (\rho / \rho_{\rm nuc})^{1/3}\, \mbox{erg K}^{-1}\mbox{cm}^{-3}$ \citep{belo16}. Here, $\dot{Q}_{B}$ and $\dot{Q}_{\nu}$ are the heating and cooling rates provided by magnetic field decay and neutrino emission, respectively. The quantity $\rho_{\text{nuc}}$ is the nuclear saturation density, which may be exceeded in the core of a particularly heavy neutron star or if the equation of state (EOS) is soft. For the \cite{akmal} EOS [which passes constraints from GW170817 \citep{gw17} and can accommodate the heaviest pulsar observed to date, PSR J0740+6620, with $M = 2.08^{+0.07}_{-0.07} M_{\odot}$ \citep{fons21}], a star of mass $1.39 M_{\odot}$ has a central density $\rho_{c} = 9 \times 10^{14}\, \mbox{g cm}^{-3} \approx 3.2 \rho_{\rm nuc}$. This increases to $\rho_{c} = 1.1 \times 10^{15}\, \mbox{g cm}^{-3}$ for a $1.66 M_{\odot}$ star.}

{Following \cite{belo16} and others, the two main cooling mechanisms we consider are the modified (mUrca) and the fast, direct Urca (dUrca) mechanisms; the former is thought to be the dominant neutrino mechanism in (non-superfluid) nucleon matter \cite[$\rho \lesssim 2 \rho_{\rm nuc}$;][]{yak04}, while the latter may activate in the core of particular dense stars \cite[$\rho \sim 4 \rho_{\rm nuc}$;][]{page91}. The presence of hyperons may reduce fast cooling thresholds \citep{anz21,anz22}.}

{The mUrca cooling rate can be approximated by \citep{max79}}
\begin{equation} \label{eq:murcaQ}
\dot{Q}_{\nu}^{M} \approx 7 \times 10^{20} \left(\frac {T_{\rm core}}{10^{9} \text{K}}\right)^{8} \left( \frac{\rho} {\rho_{\rm nuc}}\right)^{2/3} \mathcal{R}_{M} \text{ erg s}^{-1} \text{cm}^{-3},
\end{equation}
{where $\mathcal{R}_{M} \leq 1$ is a suppression factor relevant if either protons or neutrons are superfluid, whereupon the breaking of Cooper pairs instead becomes the dominant cooling mechanism at densities $\rho \sim \rho_{\rm nuc}$ \cite[e.g.,][]{page09}. We henceforth ignore such complications in our phenomenological heating model \eqref{eq:tmod}, though these should be considered in realistic magnetothermal modelling if the core temperature drops below the superfluidity onset value $ 1\lesssim T_{\rm crit}/10^{8}\text{K} \lesssim 10$ \citep{pot15}. The dUrca cooling rate is given by \citep{page91}}
\begin{equation} \label{eq:durcaQ}
\dot{Q}_{\nu}^{D} \approx 10^{27} \left(\frac{T_{\rm core}}{10^{9} \text{K}}\right)^{6} \text{ erg s}^{-1} \text{cm}^{-3},
\end{equation}
{which exceeds \eqref{eq:murcaQ} by several orders of magnitude for temperatures in the range of interest.}

{The rate of heating, provided by ambipolar diffusion, can be estimated through \citep{belo16}}
\begin{equation}
\dot{Q}_{B} \approx \frac {\tau_{\rm pn}} {\rho_{p}} \left( \frac{ B^2}{ 4 \pi L} \right)^{2},
\end{equation}
{for core field strength $B$ which varies over lengthscale $L$, where $1/\tau_{\rm pn}$ denotes the rate of p-n collisions per proton (ignoring core exotica), given by \citep{yak90}}
\begin{equation}
\tau_{\rm pn}^{-1} \approx 4.7 \times 10^{18} \left(\frac{T_{\rm core}}{10^9\text{K}}\right)^{2} \left(\frac{\rho} {\rho_{\rm nuc}}\right)^{-1/3}\, \mbox{s}^{-1}.
\end{equation}

{In the simplified model \eqref{eq:tmod}, a magnetar, born with temperature $T_{0} \lesssim 10^{11}\, \mbox{K}$, reaches a quasi-static balance temperature $T_{\rm bal}$ (i.e., $dT/dt =0$) after $\gtrsim 10$ years (even less with dUrca), where the temperature remains until field decay sets in ($\sim$kyr for $B \sim 10^{16}$G). Assuming a present-day core field of $\sim 10^{16}$G, these balance temperatures read}
\begin{equation} \label{eq:tbal}
T_{\text{bal}}^{M} \approx 8 \times 10^{8} \left( \frac {B_{16}^2} {L_{5}} \right)^{1/5} \left( \frac {\rho} {\rho_{\text{nuc}}} \right)^{-7/30} \text{ K},
\end{equation}
{for mUrca, and}
\begin{equation} \label{eq:tbalD}
T_{\text{bal}}^{D} \approx 1.3 \times 10^{8} \left( \frac {B_{16}^2} {L_{5}} \right)^{1/4} \left( \frac {\rho} {\rho_{\text{nuc}}} \right)^{-1/12} \text{ K},
\end{equation}
{for dUrca.}

Core-crust thermal transport depends primarily on the chemical composition of the stellar envelope and the magnetic stratification, which influence the photon opacity \citep{tsu72}. Given a surface temperature $T_{s}$, the flux
\begin{equation}
F_{s} = \sigma_{\text{SB}} T_{s}^4,
\end{equation}
for Stefan-Boltzmann constant $\sigma_{\text{SB}} \approx 5.67 \times 10^{-5}\text{erg cm}^{-2} \text{ s}^{-1} \text{K}^{-4}$, defines a surface luminosity
\begin{equation} \label{eq:surflum}
L_{s} = 4 \pi R_{\star}^2 \int^{1}_{0} d \left( \cos \theta \right) F_{s}.
\end{equation}
The co-latitude ($\theta$) dependence in \eqref{eq:surflum} comes through the angle between the magnetic field, assumed dipolar (see below), and the surface normal \cite[see][for more details]{belo16}, which affects the thermal conductivity tensor. For a slow source (i.e., ignoring rotational corrections to the metric tensor), the redshifted luminosity seen by an observer at infinity is then $L_{s}^{\infty} = L_{s} ( 1 - 2 G M_{\star}/c^2 \Rs)$. 

\subsection{Magneto-thermal modelling}
\label{sec:thermres}

Numerical simulations for the core-surface temperature relationship were carried out by \cite{pot03}. Using their analytic fits (which are too long to repeat here; see their Appendix A), we calculate the luminosity an observer expects to see from GLEAM-X J1627 as a function of the core field strength, assuming the system is in thermal quasi-equilibrium with balance temperature \eqref{eq:tbal} (mUrca) or \eqref{eq:tbalD} (dUrca) and that there are no other heat sources. For young ($\ll$~kyr) stars or ones where Joule heating, mechanical heating, or positron backflow from the magnetosphere is also significant, higher temperatures are expected. We further assume an iron envelope, as a crust composed of lighter elements (e.g., accreted materials) conducts heat more efficiently and predicts a higher $T_{s}$ for a given $B$. Landau quantization, which we also ignore, similarly leads to higher temperatures, because electrons are forced to move along the field lines, thereby suppressing their ability to transfer heat radially. 

\begin{figure}
\includegraphics[width=\columnwidth]{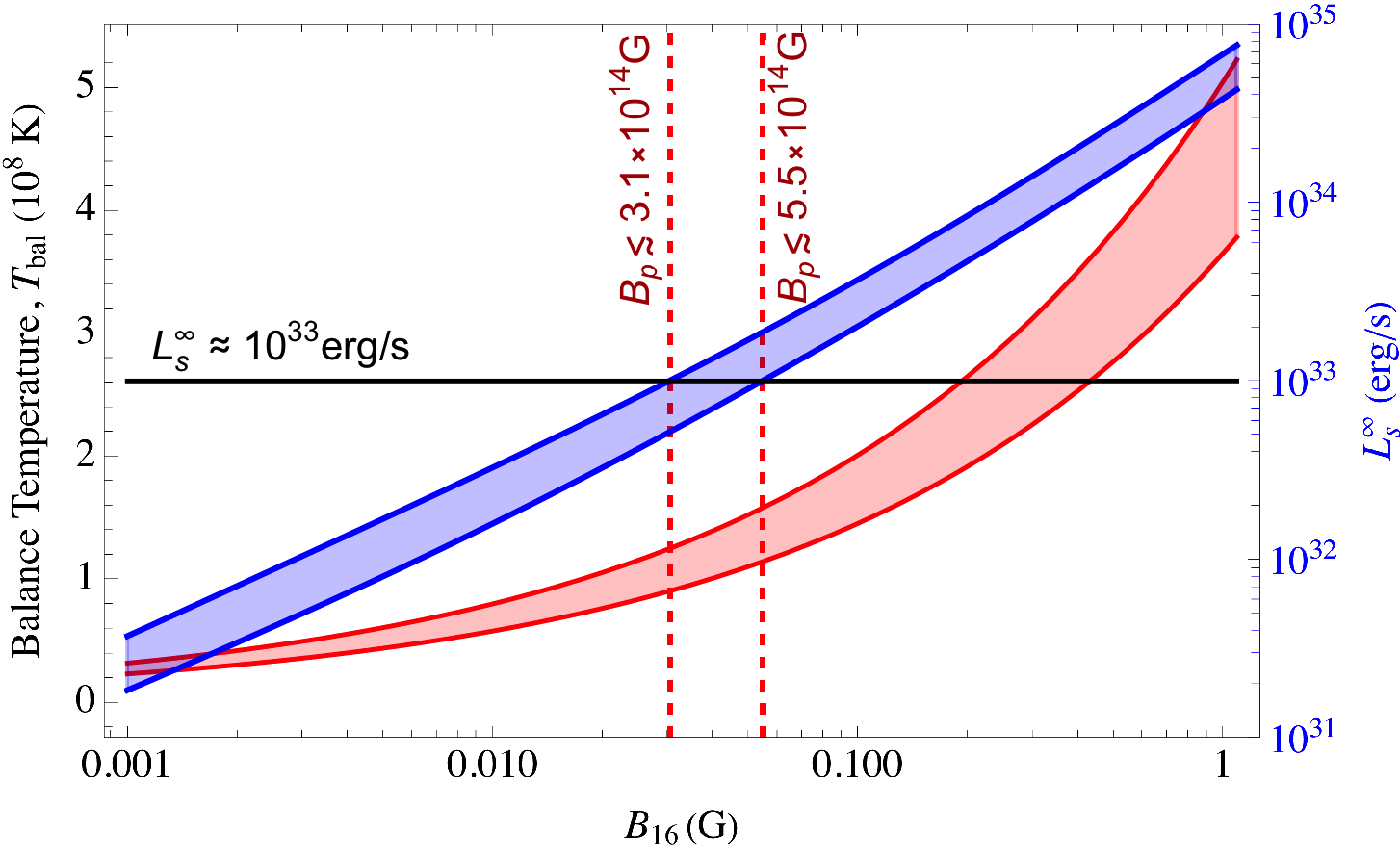}
\caption{Quasi-static core temperatures (red curves) set by balancing mUrca cooling and ambipolar heating [expression \eqref{eq:tbal}; left axis], as a function of the magnetic field strength, for two different densities, $\rho = \rho_{\text{nuc}}$ (upper curve) and $\rho = 4\rho_{\text{nuc}}$ (lower curve). The right-axis (blue curves) shows the predicted, redshift-corrected surface luminosity \eqref{eq:surflum}. An upper limit to $L^{\infty}_{s}$ implies an upper limit to the internal $B$ field, {thereby issuing a constraint on GLEAM-X J1627, for which $L_{\text{X,max}} \approx 10^{30} \text{ erg s}^{-1}$}. The solid, horizontal line corresponds to this upper limit for a conversion efficiency of $\epsilon \lesssim 0.1\%$, i.e., $L_{\text{X}} \lesssim 10^{-3} L^{\infty}_{s}$, which translates into upper limits for $B$ (dashed, vertical lines) for a given core density.}
\label{fig:xrays}
\end{figure}

\begin{figure}
\includegraphics[width=\columnwidth]{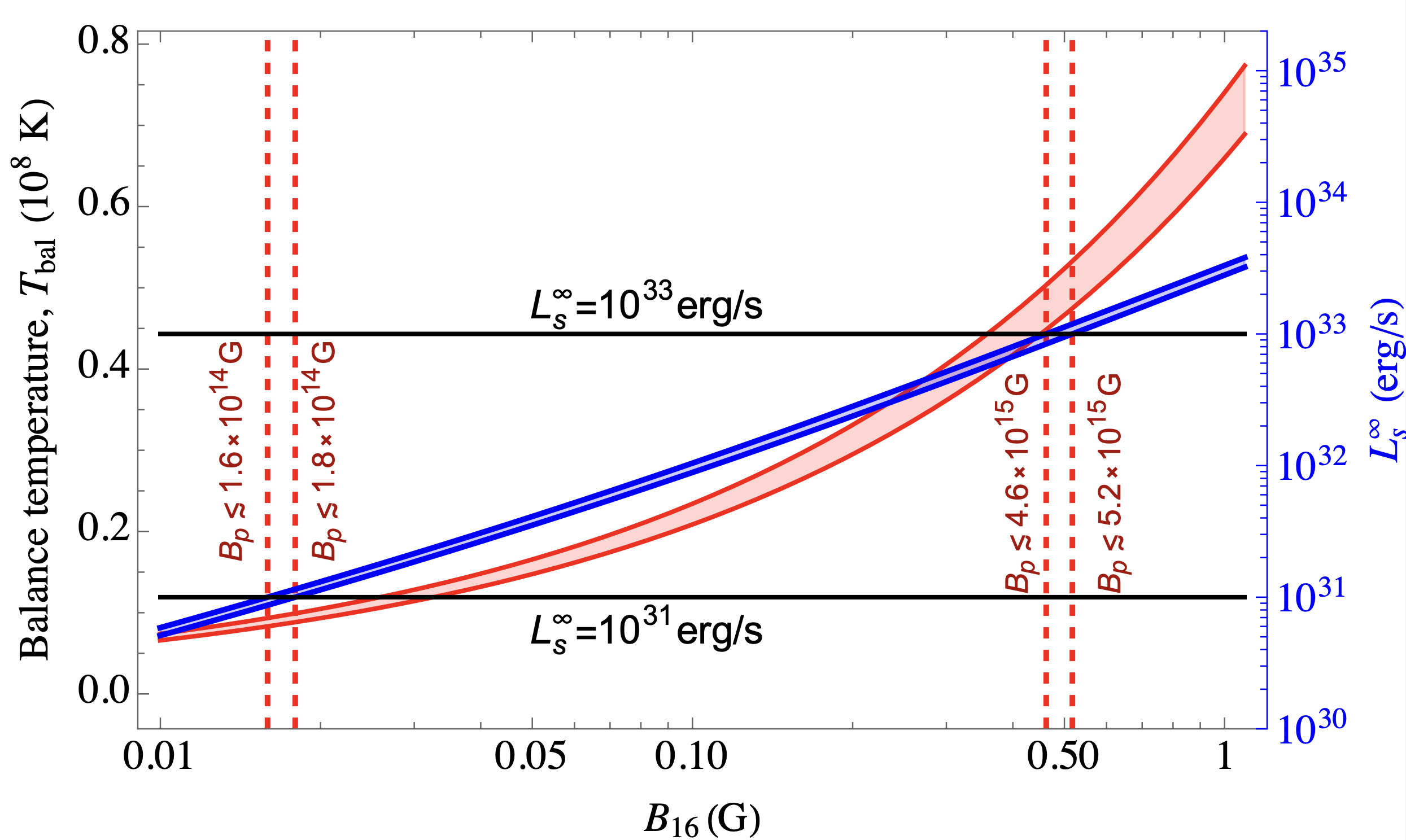}
\caption{Similar to Fig. \ref{fig:xrays} though with direct Urca cooling \eqref{eq:durcaQ} {and also a greater efficiency, $\epsilon = 10\%$ (lower, solid line).}}
\label{fig:xraysD}
\end{figure}

Figure \ref{fig:xrays} shows the balance temperature \eqref{eq:tbal} (red curves; left axis) as a function of the \emph{core} field strength, where we consider core densities of $\rho = \rho_{\text{nuc}}$ (upper curves) and $\rho = 4\rho_{\text{nuc}}$ (lower curves) {and the mUrca cooling rate \eqref{eq:murcaQ}. Figure \ref{fig:xraysD} is similar, though instead with the dUrca rate \eqref{eq:durcaQ}; note the different scales}. To provide an optimistic outlook, we take $L = \Rs$ so that the magnetic energy is predominantly concentrated in low multipoles (cf. Footnote 5).  The right {axes} (blue curves) show the surface luminosity \eqref{eq:surflum} witnessed by an observer at infinity. {These figures} illustrate that there is generally an upper limit for the core field strength implied by the absence of X-rays. For example, {even if we assume a tiny X-ray efficiency of $\epsilon \lesssim 0.1\%$ (i.e., $L_{\text{X}} \lesssim 10^{-3} L_{s}^{\infty}$), the Chandra observations of GLEAM-X J1627, which translate into an upper-limit of} $L_{\text{X}} \sim 10^{30} \text{ erg s}^{-1}$ \citep{rea22}, require core field strengths of $B \lesssim 3.1 \times 10^{14}\,\mbox{G}$ for $\rho = \rho_{\text{nuc}}$ and $B \lesssim 5.5 \times 10^{14}\,\mbox{G}$ for $\rho = 4\rho_{\text{nuc}}$, as shown by the dashed, vertical lines in Fig. \ref{fig:xrays}. {Larger, percent-level efficiencies place even tighter constraints.} The corresponding limits for dUrca are much less restrictive, viz. $B \lesssim 5.2 \times 10^{15}\,\mbox{G}$ for $\rho = 4\rho_{\text{nuc}}$ {for $\epsilon = 0.1\%$, or $B \lesssim 1.8 \times 10^{14}\,\mbox{G}$ for $\epsilon = 10\%$.}

{In the magnetothermal evolutions carried out by \cite{anz22b}, it was shown that the surface luminosity of a $1.8 M_{\odot}$ magnetar ($B \gtrsim 10^{15}\, \mbox{G}$) with Joule heating only dips below $\gtrsim 10^{33} \text{ erg s}^{-1}$ at times $t \lesssim\, \mbox{Myr}$ post-birth, even if there are hyperons and fast cooling mechanisms are active (see Figures 1 and 2 therein). This estimate, which is a factor $\sim 5$ more restrictive than the most optimistic, ambipolar model used here (see Fig. \ref{fig:xraysD}), is at odds with the minima required by the radio activation mechanisms (see Fig. \ref{fig:death}). This casts doubt on a magnetar interpretation for the source, unless the thermal luminosity is much higher, or the system is old \citep{hw22b,genc22,ben22}. We emphasise however that a realistic investigation for GLEAM-X J1627 requires a proper magnetothermal evolution in the presence of an ultra-strong field, which is difficult \cite[though see][]{rea22}.}

{We close by noting that the {magnetothermal} study of \cite{perna11} found that the waiting time distribution for flares from young ($\lesssim$\,kyr) magnetars peaks at $\sim 1$\,yr, and thus the absence of any flare phenomena in the $\sim 2$\,ks window, where the source was observed with Swift, is not entirely surprising. For a source that is several kyr old, the peak of the waiting time distribution shifts to $\gtrsim3$\,yr. Furthermore, bursts may be missed if beamed away from Earth, making it less clear how long one might need to observe before expecting a flare.} However, a plastically-flowing crust could be even hotter than one that never breaks because thermoplastic waves can dissipate magnetic energy, the effects of which resemble deflagration fronts in combustion \citep{belo16}. Targeted searches would be useful in this direction to shed light on the matter.

\section{Conclusions}
\label{sec:conc}

The source GLEAM-X J1627 was recently discovered by \cite{hw22}, who reported an extremely long spin period ($P = 1091.17$s) together with a {possibly large period derivative ($|\dot{P}| < 1.2 \times 10^{-9} \text{ ss}^{-1}$)}. The magnetic field strength implied, assuming a neutron star undergoing magnetic dipole braking \cite[though see][for a white dwarf interpretation]{loeb22,kat22}, comfortably exceeds $10^{16}\,\mbox{G}$ {when using the best-fit value $\dot{P} = 6.0 \times 10^{-10} \text{ ss}^{-1}$ \citep{rs75}}. In this paper, a critical examination of the magnetar interpretation is carried out, {though under the proviso that model-based specifics are inescapable and conclusions cannot be asserted strongly based on the limited data at hand.}

Magnetospheric gap models require a minimum magnetic field strength, for a given period, for the switch-on of the star as a radio pulsar \citep{gj69,stur71,ha01,med10}. For canonical stellar parameters, we find in Sec. \ref{sec:deathval} that minimum fields of order $\sim 10^{16}\,\mbox{G}$ appear to be necessary, even when assuming a high degree of multipolarity, {long-lived twists in the magnetosphere \citep{belo09}, and small curvature radii $R_{c} \sim R$ \citep{med10}}. If the star has a large radius, $\Rs \gtrsim 13$~km, this requirement may drop to $B_{p,\text{min}} \lesssim 5 \times 10^{15}\,\mbox{G}$. {Standard electromagnetic braking theory suggests that the star is between $\sim 10$ and $50\, \mbox{kyrs}$ old, depending on the historical braking index and field evolution model \cite[though cf.][]{hw22b,genc22,ben22}. Assuming ages much larger than $10$~kyr} and a present-day $\sim 5 \times 10^{15}\,\mbox{G}$ polar field, a Hall-Ohm back-extrapolation implies a birth strength of $\lesssim 10^{17}\,\mbox{G}$, and further that field decay was stalled to some degree, possibly by plastic opposition of the electron fluid motion in the crust \citep{lg19,g22}. This {points towards there having been a} large angular momentum reservoir at birth to support intense field amplification via some combination of dynamo activity, Kelvin-Helmholtz action, or magneto-rotational instabilities \citep{c20b,ciolfi20}. 

A simple magneto-thermal model is employed in Sec. \ref{sec:xrays} to show that the competition between heating induced by field decay and neutrino cooling implies a particular surface luminosity, depending on assumptions on the thermal conductivity, stellar composition, {and Urca channel}. {The lack of strong thermal emissions from the source \cite[$L_{\text{X}} \lesssim 10^{30}\, \mbox{erg s}^{-1}$;][]{rea22}} is difficult to reconcile with the radio requirements, unless fast cooling mechanisms are in operation \citep{page91}. A heavier star with larger moment of inertia, {which is generally easier to cool quickly \citep{anz22}}, could also help to alleviate the tension between the available spin-down power and observed radio luminosity; see Sec. \ref{sec:sec2}.

Another clue about the nature of GLEAM-X J1627 comes from the transient character of its radio pulsations. \cite{hw22} noted that the source (visibly) pulsated for only 3 months and then abruptly turned off, indicating an overall duty cycle of only $\sim 2\%$ within the observational monitoring window (see also Footnote 2). This could occur, if the source hovers near the death line, with magnetohydrodynamic evolutions triggering its descent into the graveyard around March of 2018. {For example, a crustal fracture may have injected twist into the magnetosphere prior to the object's discovery, allowing it to temporarily access line (d); see Fig. \ref{fig:death}.} In the case of the so-called rotating radio transients (RRATs), which similarly display high degrees of nulling, it was suggested by \cite{zhang07} that concentrated starspots may emerge near the poles, sporadically allowing the host star to rise above the death line \cite[see also Sec. \ref{sec:deathval},][and references therein]{suv16}. 

Magnetar-like X-ray bursts are known to suppress radio pulsations in many neutron stars; bursts observed in PSR J1119--6127 by XMM-Newton and NuSTAR were coincident with the shut-off of the source as a radio pulsar \citep{arch17}, for example {[see also \cite{coti18} for a discussion on other sources].} If GLEAM-X J1627 is regularly bursting, as would be expected if $B_{p} \gtrsim 10^{16}\,\mbox{G}$ and the crust frequently succumbs to Maxwell stresses, this could also explain the high degree of nulling. The absence of any X-ray activity \citep{hw22} casts doubt however on this interpretation, {though geometric factors related to beaming and directionality may explain this}. Finally, the population study recently conducted by \cite{she21} indicates that there is a (weak) correlation between the spin period and nulling fraction in radio pulsars. The high nulling fraction and long spin period of GLEAM-X J1627 fits within this scenario. Regardless, further monitoring of the source in both the radio and X-ray bands will help to unveil its magnetar nature or otherwise.

If indeed GLEAM-X J1627 boasts a polar field strength greater than $10^{16}\,\mbox{G}$, as {suggested} by its place in the $P$--$\dot{P}$ diagram and the death valley considerations (Sec. \ref{sec:deathval}), it would have been an ample source of gravitational waves when born. Even in the absence of a toroidal field, the quadrupolar ellipticity of the source could easily reach $\sim 10^{-4}$ \cite[e.g.,][]{hask08,mast11}. The source, located $\sim 1.3$~kpc from Earth \citep{hw22}, would have been visible to the advanced Laser Interferometer Gravitational-Wave Observatory (aLIGO) for $P \ll 1$s. From the braking analysis given in Sec. \ref{sec:brake}, if the birth period was at most a few ms (as argued in Sec. \ref{sec:birth}), the source would have been sufficiently bright in gravitational waves to enable detection for $\sim$ years. The existence of GLEAM-X J1627 therefore adds further incentive to perform blind, gravitational-wave searches for magnetar-like sources \cite[see also][]{ben22}.

\section*{Acknowledgements}
We extend our thanks to Filippo Anzuini for discussions about direct Urca processes and the heating effects of Landau quantization in magnetars. {AGS thanks Kostas Glampedakis for pointing out the possibility of a Hall attractor state.} The research leading to these results has received funding from the European Union's Horizon 2020 Programme under the AHEAD2020 project (grant n. 871158). The constructive criticisms of the anonymous referee, which led to a richer study, are gratefully acknowledged.


\section*{Data availability statement}
Observational data used in this paper are quoted from the cited works. Additional data generated from computations can be made available upon reasonable request.






\label{lastpage}


\begin{thebibliography}{96}
\expandafter\ifx\csname natexlab\endcsname\relax\def\natexlab#1{#1}\fi

\bibitem[{{Abbott} {et~al}\mbox{.}(2018){Abbott}, {Abbott}, {Abbott},
  {Acernese}, {Ackley}, {Adams}, {Adams}, {Addesso}, {Adhikari}, {Adya},
  {Affeldt}, {Agarwal}, {Agathos}, {Agatsuma}, {Aggarwal}, {Aguiar}, {Aiello},
  {Ain}, {Ajith}, {Allen}, {Allen}, {Allocca}, {Aloy}, {Altin}, {Amato},
  {Ananyeva}, {Anderson}, {Anderson}, {Angelova}, {Antier}, {Appert}, {Arai},
  {Araya}, {Areeda}, {Ar{\`e}ne}, {Arnaud}, {Arun}, {Ascenzi}, {Ashton}, {Ast},
  {Aston}, {Astone}, {Atallah}, {Aubin}, {Aufmuth}, {Aulbert}, {AultONeal},
  {Austin}, {Avila-Alvarez}, {Babak}, {Bacon}, {Badaracco}, {Bader}, {Bae},
  {Baker}, {Baldaccini}, {Ballardin}, {Ballmer}, {Banagiri}, {Barayoga},
  {Barclay}, {Barish}, {Barker}, {Barkett}, {Barnum}, {Barone}, {Barr},
  {Barsotti}, {Barsuglia}, {Barta}, {Bartlett}, {Bartos}, {Bassiri}, {Basti},
  {Batch}, {Bawaj}, {Bayley}, {Bazzan}, {B{\'e}csy}, {Beer}, {Bejger},
  {Belahcene}, {Bell}, {Beniwal}, {Bensch}, {Berger}, {Bergmann}, {Bernuzzi},
  {Bero}, {Berry}, {Bersanetti}, {Bertolini}, {Betzwieser}, {Bhandare},
  {Bilenko}, {Bilgili}, {Billingsley}, {Billman}, {Birch}, {Birney},
  {Birnholtz}, {Biscans}, {Biscoveanu}, {Bisht}, {Bitossi}, {Bizouard},
  {Blackburn}, {Blackman}, {Blair}, {Blair}, {Blair}, {Bloemen}, {Bock},
  {Bode}, {Boer}, {Boetzel}, {Bogaert}, {Bohe}, {Bondu}, {Bonilla}, {Bonnand},
  {Booker}, {Boom}, {Booth}, {Bork}, {Boschi}, {Bose}, {Bossie}, {Bossilkov},
  {Bosveld}, {Bouffanais}, {Bozzi}, {Bradaschia}, {Brady}, {Bramley},
  {Branchesi}, {Brau}, {Briant}, {Brighenti}, {Brillet}, {Brinkmann},
  {Brisson}, {Brockill}, {Brooks}, {Brown}, {Brunett}, {Buchanan}, {Buikema},
  {Bulik}, {Bulten}, {Buonanno}, {Buskulic}, {Buy}, {Byer}, {Cabero},
  {Cadonati}, {Cagnoli}, {Cahillane}, {Calder{\'o}n Bustillo}, {Callister},
  {Calloni}, {Camp}, {Canepa}, {Canizares}, {Cannon}, {Cao}, {Cao}, {Capano},
  {Capocasa}, {Carbognani}, {Caride}, {Carney}, {Carullo}, {Casanueva Diaz},
  {Casentini}, {Caudill}, {Cavagli{\`a}}, {Cavalier}, {Cavalieri}, {Cella},
  {Cepeda}, {Cerd{\'a}-Dur{\'a}n}, {Cerretani}, {Cesarini}, {Chaibi},
  {Chamberlin}, {Chan}, {Chao}, {Charlton}, {Chase}, {Chassande-Mottin},
  {Chatterjee}, {Chatziioannou}, {Cheeseboro}, {Chen}, {Chen}, {Chen}, {Cheng},
  {Chia}, {Chincarini}, {Chiummo}, {Chmiel}, {Cho}, {Cho}, {Chow},
  {Christensen}, {Chu}, {Chua}, {Chua}, {Chung}, {Chung}, {Ciani}, {Ciobanu},
  {Ciolfi}, {Cipriano}, {Cirelli}, {Cirone}, {Clara}, {Clark}, {Clearwater},
  {Cleva}, {Cocchieri}, {Coccia}, {Cohadon}, {Cohen}, {Colla}, {Collette},
  {Collins}, {Cominsky}, {Constancio}, {Conti}, {Cooper}, {Corban}, {Corbitt},
  {Cordero-Carri{\'o}n}, {Corley}, {Cornish}, {Corsi}, {Cortese}, {Costa},
  {Cotesta}, {Coughlin}, {Coughlin}, {Coulon}, {Countryman}, {Couvares},
  {Covas}, {Cowan}, {Coward}, {Cowart}, {Coyne}, {Coyne}, {Creighton},
  {Creighton}, {Cripe}, {Crowder}, {Cullen}, {Cumming}, {Cunningham}, {Cuoco},
  {Canton}, {D{\'a}lya}, {Danilishin}, {D'Antonio}, {Danzmann}, {Dasgupta}, {Da
  Silva Costa}, {Dattilo}, {Dave}, {Davier}, {Davis}, {Daw}, {Day}, {DeBra},
  {Deenadayalan}, {Degallaix}, {De Laurentis}, {Del{\'e}glise}, {Del Pozzo},
  {Demos}, {Denker}, {Dent}, {De Pietri}, {Derby}, {Dergachev}, {De Rosa}, {De
  Rossi}, {DeSalvo}, {de Varona}, {Dhurandhar}, {D{\'\i}az}, {Dietrich}, {Di
  Fiore}, {Di Giovanni}, {Di Girolamo}, {Di Lieto}, {Ding}, {Di Pace}, {Di
  Palma}, {Di Renzo}, {Dmitriev}, {Doctor}, {Dolique}, {Donovan}, {Dooley},
  {Doravari}, {Dorrington}, {Dovale {\'A}lvarez}, {Downes}, {Drago},
  {Dreissigacker}, {Driggers}, {Du}, {Dupej}, {Dwyer}, {Easter}, {Edo},
  {Edwards}, {Effler}, {Eggenstein}, {Ehrens}, {Eichholz}, {Eikenberry},
  {Eisenmann}, {Eisenstein}, {Essick}, {Estelles}, {Estevez}, {Etienne},
  {Etzel}, {Evans}, {Evans}, {Fafone}, {Fair}, {Fairhurst}, {Fan}, {Farinon},
  {Farr}, {Farr}, {Fauchon-Jones}, {Favata}, {Fays}, {Fee}, {Fehrmann},
  {Feicht}, {Fejer}, {Feng}, {Fernandez-Galiana}, {Ferrante}, {Ferreira},
  {Ferrini}, {Fidecaro}, {Fiori}, {Fiorucci}, {Fishbach}, {Fisher}, {Fishner},
  {Fitz-Axen}, {Flaminio}, {Fletcher}, {Fong}, {Font}, {Forsyth}, {Forsyth},
  {Fournier}, {Frasca}, {Frasconi}, {Frei}, {Freise}, {Frey}, {Frey},
  {Fritschel}, {Frolov}, {Fulda}, {Fyffe}, {Gabbard}, {Gadre}, {Gaebel},
  {Gair}, {Gammaitoni}, {Ganija}, {Gaonkar}, {Garcia},
  {Garc{\'\i}a-Quir{\'o}s}, {Garufi}, {Gateley}, {Gaudio}, {Gaur}, {Gayathri},
  {Gemme}, {Genin}, {Gennai}, {George}, {George}, {Gergely}, {Germain},
  {Ghonge}, {Ghosh}, {Ghosh}, {Ghosh}, {Giacomazzo}, {Giaime}, {Giardina},
  {Giazotto}, {Gill}, {Giordano}, {Glover}, {Goetz}, {Goetz}, {Goncharov},
  {Gonz{\'a}lez}, {Gonzalez Castro}, {Gopakumar}, {Gorodetsky}, {Gossan},
  {Gosselin}, {Gouaty}, {Grado}, {Graef}, {Granata}, {Grant}, {Gras}, {Gray},
  {Greco}, {Green}, {Green}, {Gretarsson}, {Groot}, {Grote}, {Grunewald},
  {Gruning}, {Guidi}, {Gulati}, {Guo}, {Gupta}, {Gupta}, {Gushwa}, {Gustafson},
  {Gustafson}, {Halim}, {Hall}, {Hall}, {Hamilton}, {Hamilton}, {Hammond},
  {Haney}, {Hanke}, {Hanks}, {Hanna}, {Hannam}, {Hannuksela}, {Hanson},
  {Hardwick}, {Harms}, {Harry}, {Harry}, {Hart}, {Haster}, {Haughian}, {Healy},
  {Heidmann}, {Heintze}, {Heitmann}, {Hello}, {Hemming}, {Hendry}, {Heng},
  {Hennig}, {Heptonstall}, {Hernandez}, {Heurs}, {Hild}, {Hinderer}, {Ho},
  {Hoak}, {Hochheim}, {Hofman}, {Holland}, {Holt}, {Holz}, {Hopkins}, {Horst},
  {Hough}, {Houston}, {Howell}, {Hreibi}, {Huerta}, {Huet}, {Hughey}, {Hulko},
  {Husa}, {Huttner}, {Huynh-Dinh}, {Iess}, {Indik}, {Ingram}, {Inta}, {Intini},
  {Irwin}, {Isa}, {Isac}, {Isi}, {Iyer}, {Izumi}, {Jacqmin}, {Jani},
  {Jaranowski}, {Johnson}, {Johnson}, {Jones}, {Jones}, {Jonker}, {Ju},
  {Junker}, {Kalaghatgi}, {Kalogera}, {Kamai}, {Kandhasamy}, {Kang}, {Kanner},
  {Kapadia}, {Karki}, {Karvinen}, {Kasprzack}, {Katolik}, {Katsanevas},
  {Katsavounidis}, {Katzman}, {Kaufer}, {Kawabe}, {Keerthana},
  {K{\'e}f{\'e}lian}, {Keitel}, {Kemball}, {Kennedy}, {Key}, {Khalili},
  {Khamesra}, {Khan}, {Khan}, {Khan}, {Khan}, {Khazanov}, {Kijbunchoo}, {Kim},
  {Kim}, {Kim}, {Kim}, {Kim}, {Kim}, {King}, {King}, {Kinley-Hanlon},
  {Kirchhoff}, {Kissel}, {Kleybolte}, {Klimenko}, {Knowles}, {Koch},
  {Koehlenbeck}, {Koley}, {Kondrashov}, {Kontos}, {Korobko}, {Korth},
  {Kowalska}, {Kozak}, {Kr{\"a}mer}, {Kringel}, {Krishnan}, {Kr{\'o}lak},
  {Kuehn}, {Kumar}, {Kumar}, {Kumar}, {Kuo}, {Kutynia}, {Kwang}, {Lackey},
  {Lai}, {Landry}, {Landry}, {Lang}, {Lange}, {Lantz}, {Lanza},
  {Lartaux-Vollard}, {Lasky}, {Laxen}, {Lazzarini}, {Lazzaro}, {Leaci},
  {Leavey}, {Lee}, {Lee}, {Lee}, {Lee}, {Lee}, {Lehmann}, {Lenon}, {Leonardi},
  {Leroy}, {Letendre}, {Levin}, {Li}, {Li}, {Li}, {Linker}, {Littenberg},
  {Liu}, {Liu}, {Lo}, {Lockerbie}, {London}, {Longo}, {Lorenzini}, {Loriette},
  {Lormand}, {Losurdo}, {Lough}, {Lousto}, {Lovelace}, {L{\"u}ck}, {Lumaca},
  {Lundgren}, {Lynch}, {Ma}, {Macas}, {Macfoy}, {Machenschalk}, {MacInnis},
  {Macleod}, {Maga{\~n}a Hernandez}, {Maga{\~n}a-Sandoval}, {Maga{\~n}a
  Zertuche}, {Magee}, {Majorana}, {Maksimovic}, {Man}, {Mandic}, {Mangano},
  {Mansell}, {Manske}, {Mantovani}, {Marchesoni}, {Marion}, {M{\'a}rka},
  {M{\'a}rka}, {Markakis}, {Markosyan}, {Markowitz}, {Maros}, {Marquina},
  {Martelli}, {Martellini}, {Martin}, {Martin}, {Martynov}, {Mason}, {Massera},
  {Masserot}, {Massinger}, {Masso-Reid}, {Mastrogiovanni}, {Matas},
  {Matichard}, {Matone}, {Mavalvala}, {Mazumder}, {McCann}, {McCarthy},
  {McClelland}, {McCormick}, {McCuller}, {McGuire}, {McIver}, {McManus},
  {McRae}, {McWilliams}, {Meacher}, {Meadors}, {Mehmet}, {Meidam},
  {Mejuto-Villa}, {Melatos}, {Mendell}, {Mendoza-Gandara}, {Mercer}, {Mereni},
  {Merilh}, {Merzougui}, {Meshkov}, {Messenger}, {Messick}, {Metzdorff},
  {Meyers}, {Miao}, {Michel}, {Middleton}, {Mikhailov}, {Milano}, {Miller},
  {Miller}, {Miller}, {Miller}, {Millhouse}, {Mills}, {Milovich-Goff},
  {Minazzoli}, {Minenkov}, {Ming}, {Mishra}, {Mitra}, {Mitrofanov},
  {Mitselmakher}, {Mittleman}, {Moffa}, {Mogushi}, {Mohan}, {Mohapatra},
  {Montani}, {Moore}, {Moraru}, {Moreno}, {Morisaki}, {Mours}, {Mow-Lowry},
  {Mueller}, {Muir}, {Mukherjee}, {Mukherjee}, {Mukherjee}, {Mukund},
  {Mullavey}, {Munch}, {Mu{\~n}iz}, {Muratore}, {Murray}, {Nagar}, {Napier},
  {Nardecchia}, {Naticchioni}, {Nayak}, {Neilson}, {Nelemans}, {Nelson},
  {Nery}, {Neunzert}, {Nevin}, {Newport}, {Ng}, {Ng}, {Nguyen}, {Nguyen},
  {Nichols}, {Nielsen}, {Nissanke}, {Nitz}, {Nocera}, {Nolting}, {North},
  {Nuttall}, {Obergaulinger}, {Oberling}, {O'Brien}, {O'Dea}, {Ogin}, {Oh},
  {Oh}, {Ohme}, {Ohta}, {Okada}, {Oliver}, {Oppermann}, {Oram}, {O'Reilly},
  {Ormiston}, {Ortega}, {O'Shaughnessy}, {Ossokine}, {Ottaway}, {Overmier},
  {Owen}, {Pace}, {Pagano}, {Page}, {Page}, {Pai}, {Pai}, {Palamos},
  {Palashov}, {Palomba}, {Pal-Singh}, {Pan}, {Pan}, {Pang}, {Pang}, {Pankow},
  {Pannarale}, {Pant}, {Paoletti}, {Paoli}, {Papa}, {Parida}, {Parker},
  {Pascucci}, {Pasqualetti}, {Passaquieti}, {Passuello}, {Patil}, {Patricelli},
  {Pearlstone}, {Pedersen}, {Pedraza}, {Pedurand}, {Pekowsky}, {Pele}, {Penn},
  {Perego}, {Perez}, {Perreca}, {Perri}, {Pfeiffer}, {Phelps}, {Phukon},
  {Piccinni}, {Pichot}, {Piergiovanni}, {Pierro}, {Pillant}, {Pinard}, {Pinto},
  {Pirello}, {Pitkin}, {Poggiani}, {Popolizio}, {Porter}, {Possenti}, {Post},
  {Powell}, {Prasad}, {Pratt}, {Pratten}, {Predoi}, {Prestegard}, {Principe},
  {Privitera}, {Prodi}, {Prokhorov}, {Puncken}, {Punturo}, {Puppo},
  {P{\"u}rrer}, {Qi}, {Quetschke}, {Quintero}, {Quitzow-James}, {Raab},
  {Rabeling}, {Radkins}, {Raffai}, {Raja}, {Rajan}, {Rajbhandari}, {Rakhmanov},
  {Ramirez}, {Ramos-Buades}, {Rana}, {Rapagnani}, {Raymond}, {Razzano}, {Read},
  {Regimbau}, {Rei}, {Reid}, {Reitze}, {Ren}, {Ricci}, {Ricker},
  {Riemenschneider}, {Riles}, {Rizzo}, {Robertson}, {Robie}, {Robinet},
  {Robson}, {Rocchi}, {Rolland}, {Rollins}, {Roma}, {Romano}, {Romel}, {Romie},
  {Rosi{\'n}ska}, {Ross}, {Rowan}, {R{\"u}diger}, {Ruggi}, {Rutins}, {Ryan},
  {Sachdev}, {Sadecki}, {Sakellariadou}, {Salconi}, {Saleem}, {Salemi},
  {Samajdar}, {Sammut}, {Sampson}, {Sanchez}, {Sanchez}, {Sanchis-Gual},
  {Sandberg}, {Sanders}, {Sarin}, {Sassolas}, {Sathyaprakash}, {Saulson},
  {Sauter}, {Savage}, {Sawadsky}, {Schale}, {Scheel}, {Scheuer}, {Schmidt},
  {Schnabel}, {Schofield}, {Sch{\"o}nbeck}, {Schreiber}, {Schuette}, {Schulte},
  {Schutz}, {Schwalbe}, {Scott}, {Scott}, {Seidel}, {Sellers}, {Sengupta},
  {Sentenac}, {Sequino}, {Sergeev}, {Setyawati}, {Shaddock}, {Shaffer}, {Shah},
  {Shahriar}, {Shaner}, {Shao}, {Shapiro}, {Shawhan}, {Shen}, {Shoemaker},
  {Shoemaker}, {Siellez}, {Siemens}, {Sieniawska}, {Sigg}, {Silva}, {Singer},
  {Singh}, {Singhal}, {Sintes}, {Slagmolen}, {Slaven-Blair}, {Smith}, {Smith},
  {Smith}, {Somala}, {Son}, {Sorazu}, {Sorrentino}, {Souradeep}, {Spencer},
  {Srivastava}, {Staats}, {Steinke}, {Steinlechner}, {Steinlechner},
  {Steinmeyer}, {Steltner}, {Stevenson}, {Stocks}, {Stone}, {Stops}, {Strain},
  {Stratta}, {Strigin}, {Strunk}, {Sturani}, {Stuver}, {Summerscales}, {Sun},
  {Sunil}, {Suresh}, {Sutton}, {Swinkels}, {Szczepa{\'n}czyk}, {Tacca}, {Tait},
  {Talbot}, {Talukder}, {Tanner}, {T{\'a}pai}, {Taracchini}, {Tasson},
  {Taylor}, {Taylor}, {Tewari}, {Theeg}, {Thies}, {Thomas}, {Thomas}, {Thomas},
  {Thorne}, {Thrane}, {Tiwari}, {Tiwari}, {Tokmakov}, {Toland}, {Tonelli},
  {Tornasi}, {Torres-Forn{\'e}}, {Torrie}, {T{\"o}yr{\"a}}, {Travasso},
  {Traylor}, {Trinastic}, {Tringali}, {Trovato}, {Trozzo}, {Tsang}, {Tse},
  {Tso}, {Tsuna}, {Tsukada}, {Tuyenbayev}, {Ueno}, {Ugolini}, {Urban}, {Usman},
  {Vahlbruch}, {Vajente}, {Valdes}, {van Bakel}, {van Beuzekom}, {van den
  Brand}, {Van Den Broeck}, {Vander-Hyde}, {van der Schaaf}, {van Heijningen},
  {van Veggel}, {Vardaro}, {Varma}, {Vass}, {Vas{\'u}th}, {Vecchio},
  {Vedovato}, {Veitch}, {Veitch}, {Venkateswara}, {Venugopalan}, {Verkindt},
  {Vetrano}, {Vicer{\'e}}, {Viets}, {Vinciguerra}, {Vine}, {Vinet}, {Vitale},
  {Vo}, {Vocca}, {Vorvick}, {Vyatchanin}, {Wade}, {Wade}, {Wade}, {Walet},
  {Walker}, {Wallace}, {Walsh}, {Wang}, {Wang}, {Wang}, {Wang}, {Wang}, {Ward},
  {Warner}, {Was}, {Watchi}, {Weaver}, {Wei}, {Weinert}, {Weinstein}, {Weiss},
  {Wellmann}, {Wen}, {Wessel}, {We{\ss}els}, {Westerweck}, {Wette}, {Whelan},
  {Whiting}, {Whittle}, {Wilken}, {Williams}, {Williams}, {Williamson},
  {Willis}, {Willke}, {Wimmer}, {Winkler}, {Wipf}, {Wittel}, {Woan}, {Woehler},
  {Wofford}, {Wong}, {Worden}, {Wright}, {Wu}, {Wysocki}, {Xiao}, {Yam},
  {Yamamoto}, {Yancey}, {Yang}, {Yap}, {Yazback}, {Yu}, {Yu}, {Yvert},
  {Zadro{\.Z}ny}, {Zanolin}, {Zelenova}, {Zendri}, {Zevin}, {Zhang}, {Zhang},
  {Zhang}, {Zhang}, {Zhang}, {Zhao}, {Zhou}, {Zhou}, {Zhu}, {Zhu}, {Zimmerman},
  {Zlochower}, {Zucker}, {Zweizig}, {LIGO Scientific Collaboration}, \& {Virgo
  Collaboration}}]{gw17}
{Abbott} B.~P. {et~al.}, 2018, \prl, 121, 161101

\bibitem[{{Aguilera}, {Pons} \& {Miralles}(2008){Aguilera}, {Pons}, \&
  {Miralles}}]{ag08}
{Aguilera} D.~N., {Pons} J.~A., {Miralles} J.~A., 2008, \apjl, 673, L167

\bibitem[{{Akg{\"u}n} {et~al}\mbox{.}(2018){Akg{\"u}n}, {Cerd{\'a}-Dur{\'a}n},
  {Miralles}, \& {Pons}}]{akg18}
{Akg{\"u}n} T., {Cerd{\'a}-Dur{\'a}n} P., {Miralles} J.~A., {Pons} J.~A., 2018,
  \mnras, 481, 5331

\bibitem[{{Akmal}, {Pandharipande} \& {Ravenhall}(1998){Akmal},
  {Pandharipande}, \& {Ravenhall}}]{akmal}
{Akmal} A., {Pandharipande} V.~R., {Ravenhall} D.~G., 1998, \prc, 58, 1804

\bibitem[{{Anzuini} \& {Melatos}(2021)}]{anz21}
{Anzuini} F., {Melatos} A., 2021, European Physical Journal A, 57, 220

\bibitem[{{Anzuini} {et~al}\mbox{.}(2022{\natexlab{a}}){Anzuini}, {Melatos},
  {Dehman}, {Vigan{\`o}}, \& {Pons}}]{anz22}
{Anzuini} F., {Melatos} A., {Dehman} C., {Vigan{\`o}} D., {Pons} J.~A.,
  2022{\natexlab{a}}, \mnras, 509, 2609

\bibitem[{{Anzuini} {et~al}\mbox{.}(2022{\natexlab{b}}){Anzuini}, {Melatos},
  {Dehman}, {Vigan{\`o}}, \& {Pons}}]{anz22b}
{Anzuini} F., {Melatos} A., {Dehman} C., {Vigan{\`o}} D., {Pons} J.~A.,
  2022{\natexlab{b}}, \mnras

\bibitem[{{Archibald} {et~al}\mbox{.}(2017){Archibald}, {Burgay}, {Lyutikov},
  {Kaspi}, {Esposito}, {Israel}, {Kerr}, {Possenti}, {Rea}, {Sarkissian},
  {Scholz}, \& {Tendulkar}}]{arch17}
{Archibald} R.~F. {et~al.}, 2017, \apjl, 849, L20

\bibitem[{{Asseo} \& {Khechinashvili}(2002)}]{ass01}
{Asseo} E., {Khechinashvili} D., 2002, \mnras, 334, 743

\bibitem[{{Baring} \& {Harding}(1998)}]{bar99}
{Baring} M.~G., {Harding} A.~K., 1998, \apjl, 507, L55

\bibitem[{{Baym} \& {Pines}(1971)}]{baym71}
{Baym} G., {Pines} D., 1971, Annals of Physics, 66, 816

\bibitem[{{Beloborodov}(2009)}]{belo09}
{Beloborodov} A.~M., 2009, \apj, 703, 1044

\bibitem[{{Beloborodov} \& {Levin}(2014)}]{bell14}
{Beloborodov} A.~M., {Levin} Y., 2014, \apjl, 794, L24

\bibitem[{{Beloborodov} \& {Li}(2016)}]{belo16}
{Beloborodov} A.~M., {Li} X., 2016, \apj, 833, 261

\bibitem[{{Beloborodov} \& {Thompson}(2007)}]{bt07}
{Beloborodov} A.~M., {Thompson} C., 2007, \apj, 657, 967

\bibitem[\protect\citeauthoryear{Beniamini et al.}{2022}]{ben22} Beniamini P., Wadiasingh Z., Hare J., Rajwade K., Younes G., van der Horst A.~J., 2022, arXiv, arXiv:2210.09323. doi:10.48550/arXiv.2210.09323


\bibitem[{{Breit} \& {Wheeler}(1934)}]{bw34}
{Breit} G., {Wheeler} J.~A., 1934, Physical Review, 46, 1087

\bibitem[{{Chen} \& {Ruderman}(1993)}]{cr93}
{Chen} K., {Ruderman} M., 1993, \apj, 402, 264

\bibitem[{{Ciolfi}(2020{\natexlab{a}})}]{c20b}
{Ciolfi} R., 2020{\natexlab{a}}, \mnras, 495, L66

\bibitem[{{Ciolfi}(2020{\natexlab{b}})}]{ciolfi20}
{Ciolfi} R., 2020{\natexlab{b}}, General Relativity and Gravitation, 52, 59

\bibitem[{{Contopoulos} \& {Spitkovsky}(2006)}]{cont06}
{Contopoulos} I., {Spitkovsky} A., 2006, \apj, 643, 1139

\bibitem[{{Coti Zelati} {et~al}\mbox{.}(2018){Coti Zelati}, {Rea}, {Pons},
  {Campana}, \& {Esposito}}]{coti18}
{Coti Zelati} F., {Rea} N., {Pons} J.~A., {Campana} S., {Esposito} P., 2018,
  \mnras, 474, 961

\bibitem[{{Duncan}(1998)}]{duncan98}
{Duncan} R.~C., 1998, \apjl, 498, L45

\bibitem[{{Eie} {et~al}\mbox{.}(2021){Eie}, {Terasawa}, {Akahori}, {Oyama},
  {Hirota}, {Yonekura}, {Enoto}, {Sekido}, {Takefuji}, {Misawa}, {Tsuchiya},
  {Kisaka}, {Aoki}, \& {Honma}}]{eie21}
{Eie} S. {et~al.}, 2021, \pasj, 73, 1563

\bibitem[{{Ek{\c{s}}i} \& {{\c{S}}a{\c{s}}maz}(2022)}]{eks22}
{Ek{\c{s}}i} K.~Y., {{\c{S}}a{\c{s}}maz} S., 2022, arXiv e-prints,
  arXiv:2202.05160

\bibitem[{{Erkut}(2022)}]{erk22}
{Erkut} M.~H., 2022, \mnras, 514, L41

\bibitem[{{Fonseca} {et~al}\mbox{.}(2021){Fonseca}, {Cromartie}, {Pennucci},
  {Ray}, {Kirichenko}, {Ransom}, {Demorest}, {Stairs}, {Arzoumanian},
  {Guillemot}, {Parthasarathy}, {Kerr}, {Cognard}, {Baker}, {Blumer}, {Brook},
  {DeCesar}, {Dolch}, {Dong}, {Ferrara}, {Fiore}, {Garver-Daniels}, {Good},
  {Jennings}, {Jones}, {Kaspi}, {Lam}, {Lorimer}, {Luo}, {McEwen}, {McKee},
  {McLaughlin}, {McMann}, {Meyers}, {Naidu}, {Ng}, {Nice}, {Pol}, {Radovan},
  {Shapiro-Albert}, {Tan}, {Tendulkar}, {Swiggum}, {Wahl}, \& {Zhu}}]{fons21}
{Fonseca} E. {et~al.}, 2021, \apjl, 915, L12

\bibitem[{{Friman} \& {Maxwell}(1979)}]{max79}
{Friman} B.~L., {Maxwell} O.~V., 1979, \apj, 232, 541

\bibitem[{{Gen{\c{c}}ali}, {Ertan} \& {Alpar}(2022){Gen{\c{c}}ali}, {Ertan}, \&
  {Alpar}}]{genc22}
{Gen{\c{c}}ali} A.~A., {Ertan} {\"U}., {Alpar} M.~A., 2022, \mnras, 513, L68

\bibitem[{{Glampedakis} \& {Lasky}(2015)}]{glamp15}
{Glampedakis} K., {Lasky} P.~D., 2015, \mnras, 450, 1638

\bibitem[{{Goldreich} \& {Julian}(1969)}]{gj69}
{Goldreich} P., {Julian} W.~H., 1969, \apj, 157, 869

\bibitem[{{Goldreich} \& {Reisenegger}(1992)}]{gr92}
{Goldreich} P., {Reisenegger} A., 1992, \apj, 395, 250

\bibitem[{{Gourgouliatos} \& {Cumming}(2014)}]{gour14}
{Gourgouliatos} K.~N., {Cumming} A., 2014, \prl, 112, 171101

\bibitem[{Gourgouliatos, De~Grandis \& Igoshev(2022)Gourgouliatos, De~Grandis,
  \& Igoshev}]{g22}
Gourgouliatos K.~N., De~Grandis D., Igoshev A., 2022, Symmetry, 14, 130

\bibitem[{{Gourgouliatos} \& {Lander}(2021)}]{gl21}
{Gourgouliatos} K.~N., {Lander} S.~K., 2021, \mnras, 506, 3578

\bibitem[{{G{\"o}{\v{g}}{\"u}{\c{s}}}
  {et~al}\mbox{.}(2000){G{\"o}{\v{g}}{\"u}{\c{s}}}, {Woods}, {Kouveliotou},
  {van Paradijs}, {Briggs}, {Duncan}, \& {Thompson}}]{gog00}
{G{\"o}{\v{g}}{\"u}{\c{s}}} E., {Woods} P.~M., {Kouveliotou} C., {van Paradijs}
  J., {Briggs} M.~S., {Duncan} R.~C., {Thompson} C., 2000, \apjl, 532, L121

\bibitem[{{Guilet} {et~al}\mbox{.}(2017){Guilet}, {M{\"u}ller}, {Janka},
  {Rembiasz}, {Obergaulinger}, {Cerd{\'a}-Dur{\'a}n}, \& {Aloy}}]{jank17}
{Guilet} J., {M{\"u}ller} E., {Janka} H.-T., {Rembiasz} T., {Obergaulinger} M.,
  {Cerd{\'a}-Dur{\'a}n} P., {Aloy} M.-A., 2017, in Supernova 1987A:30 years
  later - Cosmic Rays and Nuclei from Supernovae and their Aftermaths,
  {Marcowith} A., {Renaud} M., {Dubner} G., {Ray} A., {Bykov} A., eds., Vol.
  331, pp. 119--124

\bibitem[{{Harding}, {Contopoulos} \& {Kazanas}(1999){Harding}, {Contopoulos},
  \& {Kazanas}}]{cont99}
{Harding} A.~K., {Contopoulos} I., {Kazanas} D., 1999, \apjl, 525, L125

\bibitem[{{Haskell} {et~al}\mbox{.}(2008){Haskell}, {Samuelsson},
  {Glampedakis}, \& {Andersson}}]{hask08}
{Haskell} B., {Samuelsson} L., {Glampedakis} K., {Andersson} N., 2008, \mnras,
  385, 531

\bibitem[{{Hibschman} \& {Arons}(2001)}]{ha01}
{Hibschman} J.~A., {Arons} J., 2001, \apj, 554, 624

\bibitem[{{Ho}, {Glampedakis} \& {Andersson}(2012){Ho}, {Glampedakis}, \&
  {Andersson}}]{ho12}
{Ho} W. C.~G., {Glampedakis} K., {Andersson} N., 2012, \mnras, 422, 2632

\bibitem[{Hurley-Walker {et~al}\mbox{.}(2022)Hurley-Walker, Zhang, Bahramian,
  McSweeney, O’Doherty, Hancock, Morgan, Anderson, Heald, \& Galvin}]{hw22}
Hurley-Walker N. {et~al.}, 2022, Nature, 601, 526

\bibitem[{{Jones}(2022)}]{jones22}
{Jones} P.~B., 2022, \mnras, 510, 34

\bibitem[\protect\citeauthoryear{Katz}{2022}]{kat22} Katz J.~I., 2022, Ap\&SS, 367, 108 

\bibitem[{{Kiuchi} {et~al}\mbox{.}(2018){Kiuchi}, {Kyutoku}, {Sekiguchi}, \&
  {Shibata}}]{kiu18}
{Kiuchi} K., {Kyutoku} K., {Sekiguchi} Y., {Shibata} M., 2018, \prd, 97, 124039

\bibitem[{{Kojima}, {Kisaka} \& {Fujisawa}(2022){Kojima}, {Kisaka}, \&
  {Fujisawa}}]{koj22}
{Kojima} Y., {Kisaka} S., {Fujisawa} K., 2022, \mnras

\bibitem[{{Lander}(2022)}]{lan22}
{Lander} S.~K., 2022, arXiv e-prints, arXiv:2209.08598

\bibitem[{{Lander} {et~al}\mbox{.}(2015){Lander}, {Andersson}, {Antonopoulou},
  \& {Watts}}]{lan15}
{Lander} S.~K., {Andersson} N., {Antonopoulou} D., {Watts} A.~L., 2015, \mnras,
  449, 2047

\bibitem[{{Lander} \& {Gourgouliatos}(2019)}]{lg19}
{Lander} S.~K., {Gourgouliatos} K.~N., 2019, \mnras, 486, 4130

\bibitem[{{Lander} {et~al}\mbox{.}(2021){Lander}, {Haensel}, {Haskell},
  {Zdunik}, \& {Fortin}}]{hask21}
{Lander} S.~K., {Haensel} P., {Haskell} B., {Zdunik} J.~L., {Fortin} M., 2021,
  \mnras, 503, 875

\bibitem[{{Lander} \& {Jones}(2018)}]{land18}
{Lander} S.~K., {Jones} D.~I., 2018, \mnras, 481, 4169

\bibitem[{{Lattimer} {et~al}\mbox{.}(1991){Lattimer}, {Pethick}, {Prakash}, \&
  {Haensel}}]{page91}
{Lattimer} J.~M., {Pethick} C.~J., {Prakash} M., {Haensel} P., 1991, \prl, 66,
  2701

\bibitem[{{Lattimer} \& {Prakash}(2001)}]{lp01}
{Lattimer} J.~M., {Prakash} M., 2001, \apj, 550, 426

\bibitem[{{Loeb} \& {Maoz}(2022)}]{loeb22}
{Loeb} A., {Maoz} D., 2022, Research Notes of the American Astronomical
  Society, 6, 27

\bibitem[{{Manchester} {et~al}\mbox{.}(2005){Manchester}, {Hobbs}, {Teoh}, \&
  {Hobbs}}]{atnf}
{Manchester} R.~N., {Hobbs} G.~B., {Teoh} A., {Hobbs} M., 2005, \aj, 129, 1993

\bibitem[{{Manchester} \& {Taylor}(1977)}]{man77}
{Manchester} R.~N., {Taylor} J.~H., 1977, {Pulsars}

\bibitem[{{Mastrano} {et~al}\mbox{.}(2011){Mastrano}, {Melatos}, {Reisenegger},
  \& {Akg{\"u}n}}]{mast11}
{Mastrano} A., {Melatos} A., {Reisenegger} A., {Akg{\"u}n} T., 2011, \mnras,
  417, 2288

\bibitem[{{Medin} \& {Lai}(2010)}]{med10}
{Medin} Z., {Lai} D., 2010, \mnras, 406, 1379

\bibitem[{{Melatos}(1997)}]{mel97}
{Melatos} A., 1997, \mnras, 288, 1049

\bibitem[{{Melatos}(1999)}]{mel99}
{Melatos} A., 1999, \apjl, 519, L77

\bibitem[{{Melatos} \& {Priymak}(2014)}]{mel14}
{Melatos} A., {Priymak} M., 2014, \apj, 794, 170

\bibitem[{{Melrose}, {Rafat} \& {Mastrano}(2021){Melrose}, {Rafat}, \&
  {Mastrano}}]{melrose21}
{Melrose} D.~B., {Rafat} M.~Z., {Mastrano} A., 2021, \mnras, 500, 4530

\bibitem[{{Morozova}, {Ahmedov} \& {Zanotti}(2012){Morozova}, {Ahmedov}, \&
  {Zanotti}}]{mor12}
{Morozova} V.~S., {Ahmedov} B.~J., {Zanotti} O., 2012, \mnras, 419, 2147

\bibitem[{{Olausen} \& {Kaspi}(2014)}]{mcg14}
{Olausen} S.~A., {Kaspi} V.~M., 2014, \apjs, 212, 6

\bibitem[{{Page} {et~al}\mbox{.}(2009){Page}, {Lattimer}, {Prakash}, \&
  {Steiner}}]{page09}
{Page} D., {Lattimer} J.~M., {Prakash} M., {Steiner} A.~W., 2009, \apj, 707,
  1131

\bibitem[{{Parfrey}, {Beloborodov} \& {Hui}(2013){Parfrey}, {Beloborodov}, \&
  {Hui}}]{parf13}
{Parfrey} K., {Beloborodov} A.~M., {Hui} L., 2013, \apj, 774, 92

\bibitem[{{Perna} \& {Pons}(2011)}]{perna11}
{Perna} R., {Pons} J.~A., 2011, \apjl, 727, L51

\bibitem[{{P{\'e}tri}(2022)}]{pet22}
{P{\'e}tri} J., 2022, \aap, 657, A73

\bibitem[{{Philippov}, {Tchekhovskoy} \& {Li}(2014){Philippov}, {Tchekhovskoy},
  \& {Li}}]{phil14}
{Philippov} A., {Tchekhovskoy} A., {Li} J.~G., 2014, \mnras, 441, 1879

\bibitem[{{Potekhin}, {Pons} \& {Page}(2015){Potekhin}, {Pons}, \&
  {Page}}]{pot15}
{Potekhin} A.~Y., {Pons} J.~A., {Page} D., 2015, Space Sci. Rev, 191, 239

\bibitem[{{Potekhin} {et~al}\mbox{.}(2003){Potekhin}, {Yakovlev}, {Chabrier},
  \& {Gnedin}}]{pot03}
{Potekhin} A.~Y., {Yakovlev} D.~G., {Chabrier} G., {Gnedin} O.~Y., 2003, \apj,
  594, 404

\bibitem[{{Rankin}(1990)}]{rank90}
{Rankin} J.~M., 1990, \apj, 352, 247

\bibitem[{{Raynaud} {et~al}\mbox{.}(2020){Raynaud}, {Guilet}, {Janka}, \&
  {Gastine}}]{ray20}
{Raynaud} R., {Guilet} J., {Janka} H.-T., {Gastine} T., 2020, Science Advances,
  6, 2732

\bibitem[\protect\citeauthoryear{Rea et al.}{2022}]{rea22} Rea N., Coti Zelati F., Dehman C., Hurley-Walker N., de Martino D., Bahramian A., Buckley D.~A.~H., et al., 2022, ApJ, 940, 72 


\bibitem[{{Rea} {et~al}\mbox{.}(2007){Rea}, {Nichelli}, {Israel}, {Perna},
  {Oosterbroek}, {Parmar}, {Turolla}, {Campana}, {Stella}, {Zane}, \&
  {Angelini}}]{rea07}
{Rea} N. {et~al.}, 2007, \mnras, 381, 293

\bibitem[{{Ronchi} {et~al}\mbox{.}(2022){Ronchi}, {Rea}, {Graber}, \&
  {Hurley-Walker}}]{hw22b}
{Ronchi} M., {Rea} N., {Graber} V., {Hurley-Walker} N., 2022, \apj, 934, 184

\bibitem[{{Ruderman} \& {Sutherland}(1975)}]{rs75}
{Ruderman} M.~A., {Sutherland} P.~G., 1975, \apj, 196, 51

\bibitem[{{Sheikh} \& {MacDonald}(2021)}]{she21}
{Sheikh} S.~Z., {MacDonald} M.~G., 2021, \mnras, 502, 4669

\bibitem[{{Shibata}, {Fujibayashi} \& {Sekiguchi}(2021){Shibata},
  {Fujibayashi}, \& {Sekiguchi}}]{shib21}
{Shibata} M., {Fujibayashi} S., {Sekiguchi} Y., 2021, \prd, 103, 043022

\bibitem[{{Spitkovsky}(2006)}]{spit06}
{Spitkovsky} A., 2006, \apjl, 648, L51

\bibitem[{{Sturrock}(1971)}]{stur71}
{Sturrock} P.~A., 1971, \apj, 164, 529

\bibitem[{{Suvorov} \& {Glampedakis}(2022)}]{suvg22}
{Suvorov} A.~G., {Glampedakis} K., 2022, \prd, 105, L061302

\bibitem[{{Suvorov} \& {Kokkotas}(2019)}]{suvk19}
{Suvorov} A.~G., {Kokkotas} K.~D., 2019, \mnras, 488, 5887

\bibitem[{{Suvorov}, {Mastrano} \& {Geppert}(2016){Suvorov}, {Mastrano}, \&
  {Geppert}}]{suv16}
{Suvorov} A.~G., {Mastrano} A., {Geppert} U., 2016, \mnras, 459, 3407

\bibitem[{{Suvorov} \& {Melatos}(2020)}]{suvm20}
{Suvorov} A.~G., {Melatos} A., 2020, \mnras, 499, 3243

\bibitem[{{Szary}, {Melikidze} \& {Gil}(2015){Szary}, {Melikidze}, \&
  {Gil}}]{sz15}
{Szary} A., {Melikidze} G.~I., {Gil} J., 2015, \apj, 800, 76

\bibitem[{{Szary} {et~al}\mbox{.}(2014){Szary}, {Zhang}, {Melikidze}, {Gil}, \&
  {Xu}}]{sz14}
{Szary} A., {Zhang} B., {Melikidze} G.~I., {Gil} J., {Xu} R.-X., 2014, \apj,
  784, 59

\bibitem[{{Thompson} \& {Duncan}(1993)}]{td93}
{Thompson} C., {Duncan} R.~C., 1993, \apj, 408, 194

\bibitem[{{Thompson} \& {Duncan}(1996)}]{td96}
{Thompson} C., {Duncan} R.~C., 1996, \apj, 473, 322

\bibitem[{{Thompson} {et~al}\mbox{.}(2000){Thompson}, {Duncan}, {Woods},
  {Kouveliotou}, {Finger}, \& {van Paradijs}}]{thomp00}
{Thompson} C., {Duncan} R.~C., {Woods} P.~M., {Kouveliotou} C., {Finger} M.~H.,
  {van Paradijs} J., 2000, \apj, 543, 340

\bibitem[{{Tsuruta} {et~al}\mbox{.}(1972){Tsuruta}, {Canuto}, {Lodenquai}, \&
  {Ruderman}}]{tsu72}
{Tsuruta} S., {Canuto} V., {Lodenquai} J., {Ruderman} M., 1972, \apj, 176, 739

\bibitem[{{Turolla} {et~al}\mbox{.}(2011){Turolla}, {Zane}, {Pons}, {Esposito},
  \& {Rea}}]{tur11}
{Turolla} R., {Zane} S., {Pons} J.~A., {Esposito} P., {Rea} N., 2011, \apj,
  740, 105

\bibitem[{{Vigan{\`o}} {et~al}\mbox{.}(2013){Vigan{\`o}}, {Rea}, {Pons},
  {Perna}, {Aguilera}, \& {Miralles}}]{vig13}
{Vigan{\`o}} D., {Rea} N., {Pons} J.~A., {Perna} R., {Aguilera} D.~N.,
  {Miralles} J.~A., 2013, \mnras, 434, 123

\bibitem[{{Wadiasingh} {et~al}\mbox{.}(2020){Wadiasingh}, {Beniamini},
  {Timokhin}, {Baring}, {van der Horst}, {Harding}, \& {Kazanas}}]{wad19}
{Wadiasingh} Z., {Beniamini} P., {Timokhin} A., {Baring} M.~G., {van der Horst}
  A.~J., {Harding} A.~K., {Kazanas} D., 2020, \apj, 891, 82

\bibitem[{{Yakovlev} {et~al}\mbox{.}(2002){Yakovlev}, {Gnedin}, {Kaminker}, \&
  {Potekhin}}]{yak04}
{Yakovlev} D., {Gnedin} O., {Kaminker} A., {Potekhin} A., 2002, in 34th COSPAR
  Scientific Assembly, Vol.~34, p. 1482

\bibitem[{{Yakovlev} \& {Shalybkov}(1990)}]{yak90}
{Yakovlev} D.~G., {Shalybkov} D.~A., 1990, Soviet Astronomy Letters, 16, 86

\bibitem[{{Zhang}, {Gil} \& {Dyks}(2007){Zhang}, {Gil}, \& {Dyks}}]{zhang07}
{Zhang} B., {Gil} J., {Dyks} J., 2007, \mnras, 374, 1103

\end{thebibliography}
\end{document}